\documentclass{aa}  

\usepackage{graphicx}
\usepackage{txfonts}
\usepackage{xcolor}
\usepackage{soul}

\DeclareRobustCommand{\VAN}[3]{#2}
\let\VANthebibliography\thebibliography
\def\thebibliography{\DeclareRobustCommand{\VAN}[3]{##3}\VANthebibliography}

\newcommand{\msun}{\mathrm{M}_\odot}

\newcommand{\sextractor}{\textsc{SExtractor }}
\newcommand{\modtext}{\textcolor{black}}
\newcommand{\modtexttwo}{\textcolor{black}}

\newcommand{\magr}{M_{\mathrm{r}}}
\usepackage[]{hyperref}
\begin{document}

   \title{The intragroup light in KiDS+GAMA groups}

   \subtitle{A stacking analysis}

   \author{S.~L.~Ahad
          \inst{1, 2, 3, 4}\fnmsep\thanks{E-mail: ahad@strw.leidenuniv.nl}
          \and
          H.~Hoekstra\inst{1}
          \and
          Y.~M.~Bah\'{e}\inst{5,6}
          \and
          I.~K.~Baldry\inst{7}
          \and
          K.~Kuijken\inst{1}
          \and
          S.~Brough\inst{8}
          \and 
          B.~W.~Holwerda\inst{9}}

   \institute{Leiden Observatory, Leiden University, P.O. Box 9513, 2300 RA Leiden, The Netherlands
        \and
        Waterloo Centre for Astrophysics, University of Waterloo, Waterloo, ON N2L 3G1, Canada
        \and
        Department of Physics and Astronomy, University of Waterloo, Waterloo, ON N2L 3G1, Canada
        \and
        Center for Astronomy, Space Science, and Astrophysics, Independent University, Dhaka 1229, Bangladesh
        \and
        School of Physics and Astronomy, University of Nottingham, University Park, Nottingham NG7 2RD, UK
        \and
        Laboratory of Astrophysics, Ecole Polytechnique F\'{e}d\'{e}rale de Lausanne (EPFL), Observatoire de Sauverny, 1290 Versoix, Switzerland
        \and 
        Astrophysics Research Institute, Liverpool John Moores University, IC2, Liverpool Science Park, 146 Brownlow Hill, Liverpool L3 5RF, UK
        \and
        School of Physics, University of New South Wales, NSW 2052, Australia
        \and
        Department of Physics and Astronomy, University of Louisville, Natural Science Building 102, Louisville KY 20494, USA
        }

   \date{Received xxxx xx, 2024; accepted xxxx xx, xxxx}

 
  \abstract{Galaxy groups and clusters assembled through dynamical interactions of smaller systems, resulting in the formation of a diffuse stellar halo known as the intragroup or intracluster light (IGL/ICL). By preserving the records of these interactions, the IGL/ICL provides valuable insight into the growth history of galaxy groups and clusters. Groups are especially interesting because they represent the link between galactic haloes and massive clusters. However, the low surface brightness of this diffuse light makes it extremely challenging to detect individually. Recent deep wide-field imaging surveys allow us to push such measurements to lower brightness limits by stacking data for large ensembles of groups, suppressing the noise and biases in the measurements. In this work, we present a special-purpose pipeline to reprocess individual $r-$band Kilo-Degree Survey (KiDS) exposures to optimise the IGL detection. Using an initial sample of 2385 groups with at least five spectroscopically-confirmed member galaxies from the Galaxy and Mass Assembly (GAMA) survey and deep images from the KiDS (reprocessed with our updated pipeline), we present the first robust measurement of IGL from a large group sample ($\sim 750$) down to 31-32 mag/arcsec$^2$ (varying in different stacked bins). We also compare our stacked IGL measurements to predictions from matched mock observations from the Hydrangea cosmological hydrodynamic simulations. Systematics in the imaging data can affect IGL measurements, even with our special-purpose pipeline. However, with a large sample and optimised analysis, we can place well-constrained upper and lower limits on the IGL fraction (3 - 21 per cent) for our group ensemble across $0.09\leq z\leq 0.27$ and $12.5\leq \log_{10}[M_{200}/\msun] \leq 14.0$. This work explores the potential performance of stacked statistical analysis of diffuse light in large samples of systems from next-generation observational programs like \textit{Euclid} and the Vera C. Rubin Observatory's Legacy Survey of Space and Time (LSST).}


   \maketitle
%

\section{Introduction}

It is well-established that the central galaxies (CG, commonly referred to as the brightest group/cluster galaxy, BGG/BCG) in groups and clusters of galaxies are surrounded by an extended diffuse distribution of stars, which are often referred to as the intragroup or intracluster light \citep[IGL/ICL, see e.g.][for recent reviews]{Mihos2015,Contini2021,Montes2022,arnaboldi2022}. Extending out to several hundreds of kilo-parsecs from the centre and often enveloping multiple galaxies in the host system, this diffuse light is generally considered as a separate component of the galaxy groups and clusters they are part of. Over the last few decades, different techniques have been explored to separate this diffuse component from their host galaxy and measure the amount of light originating from it. Depending on the measuring technique, the ICL has been found to contain as much as 30 per cent or more of the total starlight of the host system \citep[e.g.,][]{zibetti2005,Gonzalez2013,Mihos2017,Montes2018,Zhang2019,Kluge2021}. However, a unanimous definition of this diffuse component (from simulations and observations) and how much they contribute to the total light of their host systems is still an open discussion \citep[see, e.g. table 1 from][]{Kluge2021,brough2024}.  

In recent years, studies of the ICL in individual clusters using deep imaging have been increasing \citep[e.g.,][]{Mihos2005,seigar2007,Montes2014,Montes2018, JimenezTeja2018,Demaio2020,Montes2021,garatenunez2024}, along with a few works where stacking a statistical sample of systems to improve the signal-to-noise-ratio have been performed \citep[e.g.,][]{zibetti2005,Zhang2019,Zhang2024}. The origin and growth history of ICL has been explored through multiple simulation-based studies \citep[see, e.g.][for more discussion on the origin and growth of the ICL]{Mihos2017,Contini2021}. These studies found several ways in which the IGL/ICL can build up, including tidal stripping \citep{Gallaghar1972}, galaxy disruption \citep{Guo2011}, galaxy mergers \citep{Murante2007}, and in situ star formation in the intracluster medium \citep{Puchwein2010,Tonnesen2012}. Several recent works indicate that the CG+ICL distribution follows the global dark matter (DM) distribution (e.g., \citealt{Montes2019,Zhang2019,sampaioS2021} based on observations, and \citealt{AlonsoAsensio2020,yoo2024} from simulations), and that \modtext{ compared to the CG, ICL distribution on average aligns more with the underlying cluster (galaxy) distribution \citep{pillepich2014,Kluge2021}}. These findings imply that the IGL/ICL growth is connected to the build-up of their host systems, and therefore, is a great probe to understand the evolution of large-scale structures like galaxy clusters and the galaxies within such systems. 

Although most studies on this diffuse light are based on clusters because the ICL is more prominent, and clusters are preferentially targeted by deep surveys such as Hubble Frontier Fields \citep[HFF,][]{Lotz2017} or Beyond Ultra-deep Frontier Fields And Legacy Observations \citep[BUFFALO,][]{steinhardt2020}, studying the diffuse light in groups (or IGL) is particularly interesting for several reasons. Firstly, groups cover the intermediate halo mass regime between galaxy haloes and galaxy cluster haloes. Therefore, understanding the build-up of the diffuse light across a wide halo mass range of the host systems requires an understanding of the growth of IGL as well. Secondly, groups are interesting and distinct systems compared to the clusters because they are dynamically less disturbed, and have had fewer interactions with other systems. As a result, it is more straightforward to connect the growth of the IGL in groups with their dynamic history. Finally, according to the hierarchical structure formation model, the larger clusters are built by the infall and merging of smaller groups in their already existing larger haloes. In this scenario, \modtext{a good fraction of the dynamical interactions and preprocessing of the member galaxies already happened in the groups before they even became part of a larger halo or a cluster. A byproduct of the dynamical interactions of member galaxies in groups would be the presence of IGL at a cosmic epoch when most large clusters were yet to form}. Detection of IGL in a $z=1.85$ galaxy group by \citet{coogan2023} supports this scenario. Recent works on detecting ICL in high-redshift clusters and protoclusters also provide evidence of the growth of this diffuse light for a long time over the age of the Universe \citep[e.g.][]{joo2023,werner2023}. Therefore, understanding these smaller systems will lead to a better understanding of the physics of the larger systems. 

Even though the importance of understanding the fractions and buildup of IGL/ICL across a wide range of host halo mass is clear, there have only been a few studies on IGL, and even fewer that cover a wide range of group-mass haloes. The main reason behind this is the lack of deep data with high enough resolution and signal-to-noise ratio (SNR) to reliably detect and analyze the faint IGL in groups. 
Studying the light distribution of individual groups is useful to understand the diversity of the IGL signal and its formation channels \citep[e.g.][]{Demaio2020,ragusa2023,martinezlombilla2023}. However, the low surface brightness of the IGL means that individual systems have a very low SNR, which makes measurements more susceptible to systematics in the data and introduces a higher uncertainty in their interpretations. Stacking the light of multiple groups can help to improve the SNR while keeping the key features of the underlying population. 
\citet{zibetti2005} studied the diffuse light in 683 SDSS groups and clusters at $0.2<z<0.3$ using $g$, $r$, and $i$ band photometry by stacking them to increase the SNR. They reported that, on average, the ICL contributes a small fraction ($\sim$ 10\%) of the total visible light in a cluster. They also found that the surface brightness of the ICL correlates with BCG luminosity and with cluster richness, but the fraction of the total light in the ICL does not vary notably with these properties. However, they only studied these behaviours by dividing their sample into two sub-samples for each property (bright BCG - faint BCG, high richness - low richness) which may not be sensitive to a wider variation of these properties. 
With the group catalogue based on spectroscopic redshifts by the Galaxy and Mass Assembly \citep[GAMA,][]{Driver2009,driver2011} survey and deep multi-band ($u, g, r, i$) photometry of the Kilo-Degree Survey \citep[KiDS,][]{Kuijken2019} covering the same region as in the GAMA catalogue, we can now attempt to push the detection limit of the IGL and explore its co-evolution with the host systems across a wider halo mass range.

However, before simply stacking all the group data, we need to consider a few caveats. One important issue is the diversity of IGL/ICL distribution depending on the properties of the host system (groups/clusters) and its central galaxy (CG). Based on the data from their semi-analytic model, \citet{ContiniGu2021} reported that the ICL distribution varies widely depending on the dynamical history and morphology of the CG. Another recent work based on 170 low-redshift ($z \leq 0.08$) galaxy clusters in the northern hemisphere by \citet{Kluge2021} reported a positive correlation between CG+ICL brightness and different properties of the host cluster (e.g., mass, size, and integrated light in the satellites). Therefore, it is necessary to quantify the effect of galaxy and host system properties on the IGL measurements in a stacking analysis in order to find the optimal way of stacking for a reliable interpretation of the measurements. We explored this in \citet{ahad2023}, where we used mock observations of a GAMA-like group sample matching the  KiDS $u-$ and $r-$band photometry using the Hydrangea cosmological simulations \citep{bahe2017hydrangea}. We utilise insights and predictions from \citet{ahad2023} in this work to design and interpret our analysis.

Another major concern is the suitability of KiDS data for low-surface-brightness (LSB) analysis such as IGL measurement because of this being a cosmology survey with imaging from a wide-field camera. The data processing pipelines for cosmology surveys are usually optimised for measuring shapes and fluxes of small and faint galaxies. This requires a uniform photometric zero-point throughout the large joined pointings, which is often achieved by background-level detection and subtraction on very small scales compared to the total image size. The resulting images can have an uneven background, with the background over-subtracted near bright sources such as the CG of groups and clusters, making them quite unsuitable for IGL measurements \citep[e.g.][]{Furnell2021,Montes2021,martinezlombilla2023}. Moreover, in wide-field cameras like the OmegaCAM \citep{Kuijken2011}, the large aperture can cause internal reflection of light from bright sources, resulting in residual (radial) patterns in the field image from uneven illumination. A wider field of view also increases the chance of streaks of diffused stray light from bright sources that are nearby, such as the Moon, planets, or artificial satellites. The stray-light and internal reflection issues are usually taken care of during the data processing phase. However, the standard corrections can leave extremely faint residual patterns in the field-of-view that only surface while stacking many images. This should be accounted for, especially in the case of an LSB analysis. Therefore, we develop a custom-made pipeline to re-process the KiDS data, taking special care of the background subtraction to retain a uniform background as much as possible.

In this paper, we present our custom pipeline to reprocess the multi-band ($u, g, r, i$) imaging from the KiDS data release 4 to optimise them for LSB analysis, and different checks that were done to ensure a robust measurement of the faint IGL in GAMA groups. We also present IGL measurements in stacked groups of different luminosities and redshift bins and compare them with predictions from the Hydrangea cosmological hydrodynamic simulations.

The organization of the paper is as follows. In Sec.~\ref{sec:data_des}, we present the GAMA groups and KiDS multi-band data we use for this work along with  the selection criteria for our group sample. We also discuss the necessity of a custom pipeline for our analysis and describe the pipeline and its performance in keeping a uniform background level in the data in detail in this section (from Sec.~\ref{sec:processing} onward). In Sec.~\ref{sec:galaxy_profile_and_masking_sat}, we test the performance of the pipeline on retaining the diffuse light in galaxy outskirts and explore how to mask the diffuse light of satellite galaxies from the IGL measurements. In Sec.~\ref{sec:sim_prediction}, we build an updated PSF model from the re-processed KiDS images and evaluate its impact on simulated IGL measurements. In Sec.~\ref{sec:igl_chapt}, we present our resulting measurements and discuss how they compare to our predictions from simulations. Finally, in Sec.~\ref{sec:conclusions}, we discuss the performance and expectation from wide-field surveys like KiDS in LSB analysis such as IGL measurement and summarise our findings.

A flat $\Lambda$CDM cosmology is assumed for any relevant calculations in this work, with $H_0$ = 70 kms$^{-1}$Mpc$^{-1}$, $\Omega_{\Lambda} = 0.7$, and $\Omega_\textrm{M} = 0.3$. 


\section{Data}
\label{sec:data_des}
\subsection{Galaxy and Mass Assembly survey}
\label{sec:gama_data_intro}

The Galaxy And Mass Assembly (GAMA) project is a unique galaxy survey \citep{Driver2009,driver2011} with 21-band photometric data and spectroscopic redshifts for $\sim 300,000$ galaxies. The high spectroscopic completeness of the survey (98.5\% complete at $r-$band magnitude $< 19.8$ mag for SDSS-selected galaxies, \citealt{Liske2015}) allows for an excellent group selection \citep{Robotham2011}. The galaxy spectra in the GAMA survey were primarily measured by the AAOmega multi-object spectrograph on the Anglo-Australian Telescope (AAT) in five fields covering a total of $\sim$ 286 deg$^2$ area. 
Four of the GAMA fields (equatorial G09, G12 and G15 of 60 deg$^2$ each, and Southern G23 of $\sim$ 51 deg$^2$) entirely overlap with the Kilo-Degree Survey \citep[KiDS,][]{dejong2013,dejong2015,dejong2017,Kuijken2019} -- a large, deep, multi-band optical imaging survey that has great potential to reveal the faint IGL in GAMA groups (details in the following section).

We used the latest GAMA-II Galaxy Group Catalogue \citep[\texttt{G$^3$CFoFv08},][]{Robotham2011} in this work. The catalogue was generated using a friends-of-friends (FoF) based grouping algorithm where galaxies are grouped based on their line-of-sight and projected physical separations. Information about the group member galaxies was obtained using an accompanying galaxy catalogue, \texttt{G$^3$CGalv09} \citep{Robotham2011,Liske2015}. 

\begin{figure*}
\sidecaption
\leavevmode \hbox{%
  \includegraphics[width=12cm]{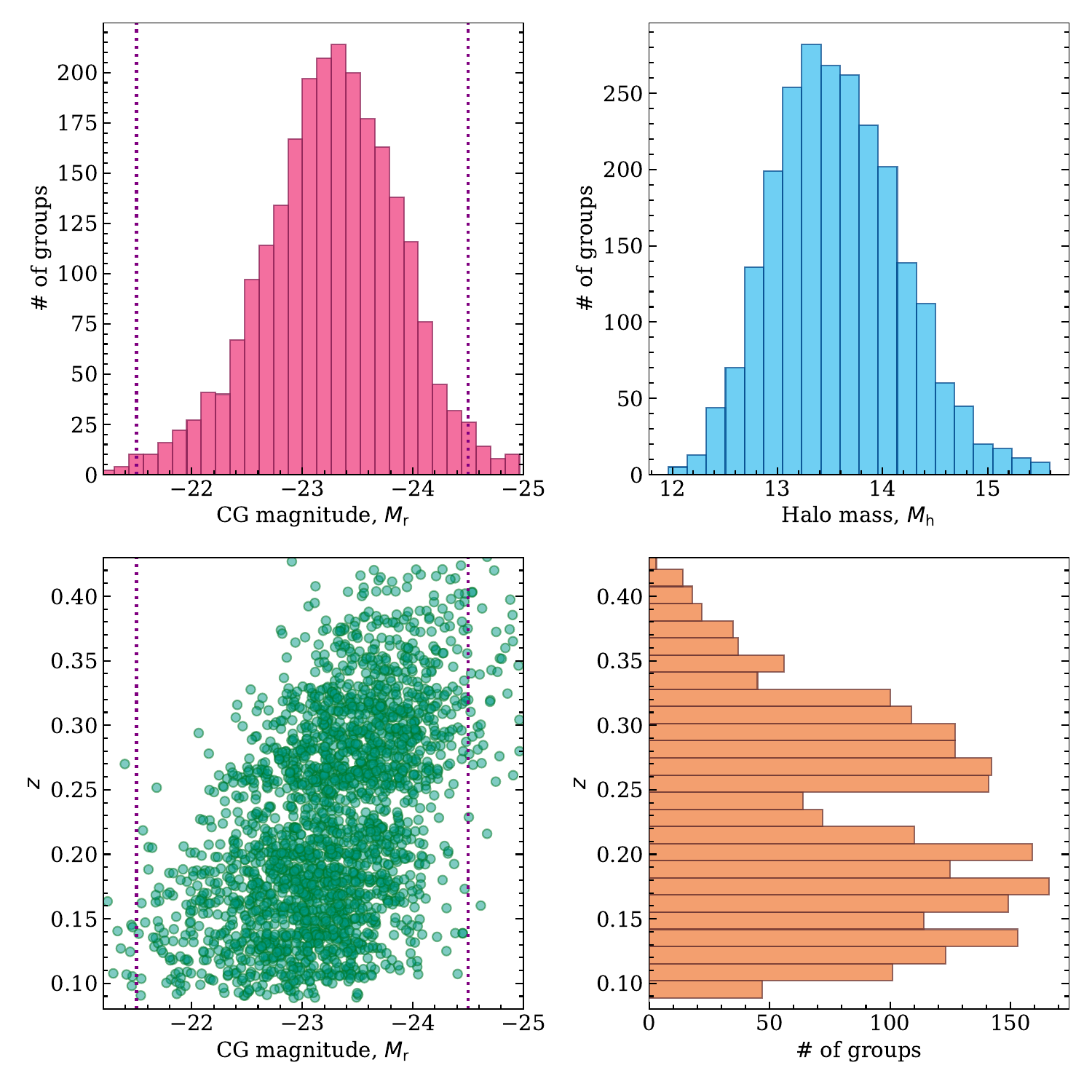}}
    \caption{Distributions of different properties of the GAMA groups with $N_{\textrm{FoF}}\geq5$ in our sample. The absolute $r-$band magnitudes ($M_{\mathrm{r}}$) and redshifts ($z$) of the central galaxies (CG) were directly obtained from the GAMA-II Galaxy Group Catalogue \citep[\texttt{G$^3$CFoFv08},][]{Robotham2011}. The halo masses were computed from the total $r-$band group luminosity using eqn.~37 of \citet{viola2015}. The vertical lines in the left panels indicate the magnitude range of the group CGs used in this work.
 \label{fig:gama_kids_groups}}
\end{figure*}

To ensure the most robust group selection, we only considered groups with 5 or more member galaxies ($N_{\textrm{FoF}}\geq5$). After applying the $N_{\textrm{FoF}}$ selection cut, we obtained a sample of 2389 groups. The distribution of redshift, CG (the iterative central galaxy from the \texttt{G$^3$CFoFv08} catalogue) magnitude, and halo mass of our final GAMA group sample is shown in Fig.~\ref{fig:gama_kids_groups}. The group halo masses were computed using the total $r-$band luminosity of the groups from the \texttt{G$^3$CFoFv08} catalogue (`LumB' parameter). We used the functional form presented by eqn.~37 of \citet{viola2015} for this halo-mass calculation, which was based on the total $r-$band luminosity to halo mass (from weak lensing measurements) scaling relation. 

We used the stellar mass estimates and $r$-band magnitudes of GAMA galaxies from the \texttt{StellarMassesLambdarv20} catalogue \citep{Taylor2011,Wright2016}. This catalogue provides physical parameters based on stellar population fits to rest-frame $ugrizY$ spectral energy distributions (SEDs), and matched aperture photometry measurements of SDSS and VIKING photometry for all the $z < 0.65$ galaxies in the GAMA-II equatorial survey regions. This sample contains over 192,000 galaxies, and the stellar mass measurements assume $H_0 = \textrm{70km s}^{-1} \textrm{Mpc}^{-1}$. Further details on the GAMA stellar mass derivation can be found in \citet{Taylor2011} and \citet{Wright2016}. 

\subsection{Kilo-Degree Survey}

The Kilo-Degree Survey \citep[KiDS,][]{dejong2013}
is a large, deep, multi-band optical imaging survey that covers 1350 square degrees in four broadband filters $(u,g,r,i)$. This cosmology survey was designed with the primary goal of mapping the large-scale matter distribution in the Universe and constraining the equation-of-state of dark energy \citep[some recent results can e.g. be found in:][]{giblin2021,li2023,burger2023}. The cosmological analysis includes measuring the effect of line-of-sight large-scale structures on galaxy shapes due to weak gravitational lensing. KiDS imaging was obtained with the square 268-million pixel CCD mosaic camera OmegaCAM \citep{Kuijken2011} that covers a $1.013^{\circ} \times 1.020^{\circ}$ area at $0\farcs 213$ pitch at the VLT Survey Telescope \citep[VST;][]{Capaccioli2011,capaccioli2012science}. The best seeing conditions (FWHM < $0\farcs 8$) were used for exposures in the $r-$band filter in order to take deep images (mean limiting $m_r = 25.02$ within $5\sigma$ in a $2''$ aperture) for the measurement of galaxy shapes. The GAMA group catalogue with spectroscopically confirmed member galaxies, accompanied by the deep KiDS imaging, provides us with a unique opportunity to analyze the IGL around the low-mass galaxy groups. 

The optical imaging of the survey included in the public data release was produced by two dedicated pipelines. The \textsc{AstroWISE} information system \citep{McFarland2013} was used for producing the co-added images in the $ugri$ bands, and the \textsc{theli} \citep{Erben2005,schirmer2013} pipeline was used to separately reduce the $r-$band images for providing a suitable source catalogue for the core weak lensing science case. These pipelines were optimized to have a uniform photometric zero-point throughout the full mosaic, which is essential for measuring shapes and photometry of small faint galaxies. However, the sky background is defined locally (by interpolating a $3\times3$ pixels median-filtered map of background estimates in $128\times128$ pixel blocks), and can be over-estimated around bright sources (like central group/cluster galaxies) in the resulting images. Such background over-subtraction does not impact the galaxy shape measurements, however, it strongly affects the faint diffuse light around bright galaxies. The final outputs from the standard KiDS pipeline are therefore unsuitable for low-surface-brightness (LSB) analyses such as a measurement of the IGL, and require a re-processing of the raw images to retain the faint light. We explain the reasoning more in Sec.~\ref{sec:processing}, and introduce our updated pipeline to reprocess the KiDS data for IGL analysis in Sec.~\ref{sec:pipeline}  (also see Fig.~\ref{fig:psf_global} for the impact of the standard and updated pipeline on PSF profile). For further details on the latest (fourth) public data release of the KiDS survey, and the image reduction procedure, we refer the interested reader to \citet{Kuijken2019}. 

\subsection{Customized data processing}
\label{sec:processing}

Large cosmological imaging surveys, such as KiDS, aim to detect small, faint galaxies and measure their positions, fluxes, and shapes. Commonly, multiple exposures (typically five in the case of KiDS) are combined to obtain a deeper image, which in turn is used for object detection. As the exposures are offset in position, and the background varies between them, an estimate for the background is subtracted before combining the data. This avoids imprinting the pattern of the individual chips in the final combined images used for object detection and photometry.

As the background also varies spatially, subtracting a constant value does not suffice. The standard KiDS pipeline uses \textsc{Swarp} \citep{bertin2002}, which estimates the background on a mesh grid. The resulting values are clipped to remove outliers that may arise from the presence of bright stars. \modtext{The mesh size ($128\times128$ pixels) sets the scale on which background variations can be captured, and it is typically chosen to be significantly smaller than the size of the chip ($\sim2000\times4000$ pixels), such that the spatial variations can be captured \citep{Kuijken2019}.} A cubic spline is then fit through the remaining samples and this background model is subtracted. As cosmological applications focus on galaxies that are much smaller than the mesh size, this approach is adequate for KiDS science goals. However, the extended and diffuse IGL can be treated as a part of the background because of their large spans and is mostly removed in this pipeline.

Moreover, the presence of bright objects can bias the background estimate locally, thus leading to overestimating the background near those locations \citep[e.g.][]{aihara2019,watkins2024}. As discussed in more detail in Sec.~\ref{sec:psf}, this leads to a region of negative flux around bright stars. Similarly, we find that the surface brightness profiles of the group central galaxies are also affected. It may be possible to, at least partially, alleviate this problem by post-processing the survey images, as was done in \cite{Furnell2021}. We take a different approach and reprocess the KiDS imaging data, attempting to avoid this issue altogether, \modtexttwo{or at least, minimize it}.

For robust IGL measurements, we need to ensure that the background estimation is not correlated with the objects of interest, \modtext{that is, the background is not estimated on the same size scale as the objects of interest}. Provided the fact that we are averaging a large number of profiles, the impact of residual flux is to increase the uncertainty in the measurements. The residuals, which may be artefacts, scattered light from stars or galaxies below the detection limit, introduce inhomogeneities in addition to the sky noise in the images \citep[e.g.][]{uson1991,slater2009,bazkiaei2024}. This is a major advantage of stacking the profiles of CGs, compared to analysing individual objects. In the latter, residuals may be difficult to distinguish from the signal of interest, whereas in a stacking analysis, residuals contribute to an \modtexttwo{almost} uniform background, albeit with increased noise. The only remaining concern is the contribution from satellite galaxies associated with the CGs, something we will explore in Sec.~\ref{sec:mask_test}. We found that varying scattered light, in the end, is a limiting factor in these data.

Although we expect a stacking approach to be more robust (as discussed above), we nonetheless wish to reduce the contribution from residuals as much as possible, to ensure that they are a subdominant contributor to the uncertainty in the measurements of the surface brightness profile.
To this end, we developed an independent pipeline for the sole purpose of measuring the low surface brightness around bright galaxies in KiDS. In Sec.~\ref{sec:pipeline} we describe the various steps in the analysis and test the performance in Sec.~\ref{sec:bgtest}. We demonstrate the value of our dedicated pipeline by measuring the average surface brightness profile around bright stars in Sec.~\ref{sec:psf}.

\subsubsection{Description of the pipeline}
\label{sec:pipeline}

For our purpose, it is not necessary to combine the exposures of a pointing before measuring the profiles. Instead, we measure the profiles around the galaxies in each exposure and average these at a later stage. In principle, creating a catalogue with object detection from a stack would allow us to identify and mask fainter galaxies, but as the images are sky-background limited, we expect this to lead to a negligible improvement, while complicating the pipeline. We therefore process individual exposures in the various filters. 

We start with the bias-subtracted and flat-fielded images (`reduced science frames') from the \textsc{AstroWISE} archive\footnote{\href{http://www.astroWISE.org}{http://www.astroWISE.org}}. This ensures that the pixel response non-uniformities are accounted for. We want to ensure that we start with images with a minimal spatial variation in the sky level and with a minimum of coherent background features. To this end, we created a new flatfield from these science exposures. To do so, we first identified objects using \textsc{SExtractor} \citep{bertin_arnouts1996} and masked those. The unmasked pixels were used to create the new flatfield, \modtexttwo{or `delta flat', similar to \citet{demaio2015}}. \modtext{This approach is commonly used in the literature, e.g. in \citet{gonzalez2005, demaio2015,kluge2020ApJS, watkins2024}. 
We used the five exposures for each of the 192 field images that contains one or more GAMA groups in our sample to create this combined delta flat. Ideally, the resulting flatfield would result in images with a constant background, but unfortunately, this turned out not to be the case.}

\begin{figure}
\centering
\leavevmode \hbox{
  \includegraphics[width=\columnwidth]{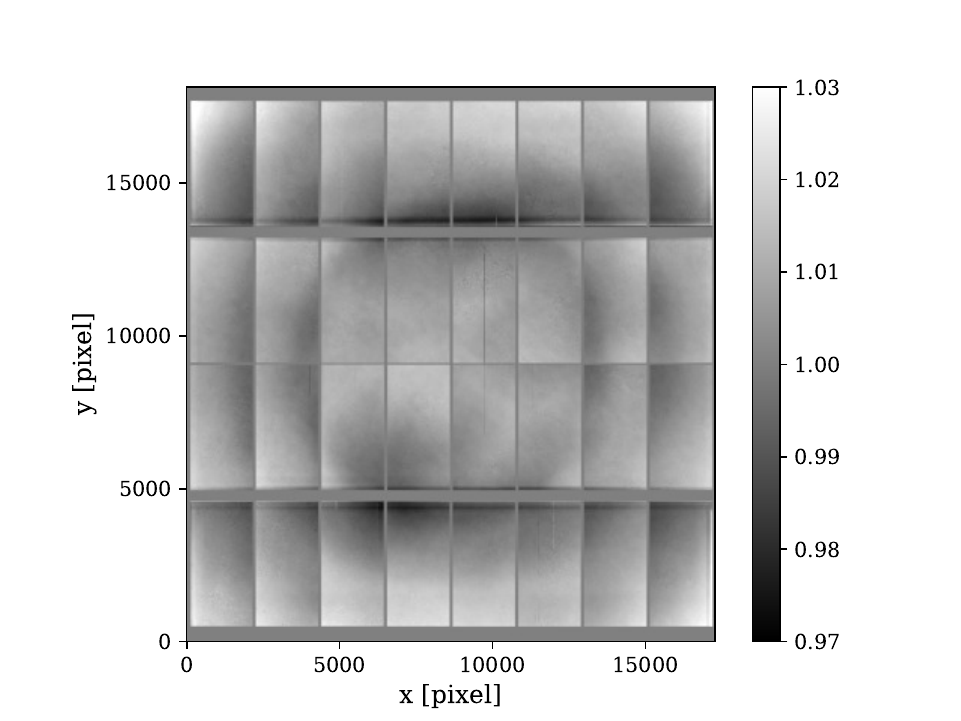}}
   \includegraphics[width=8.5cm]{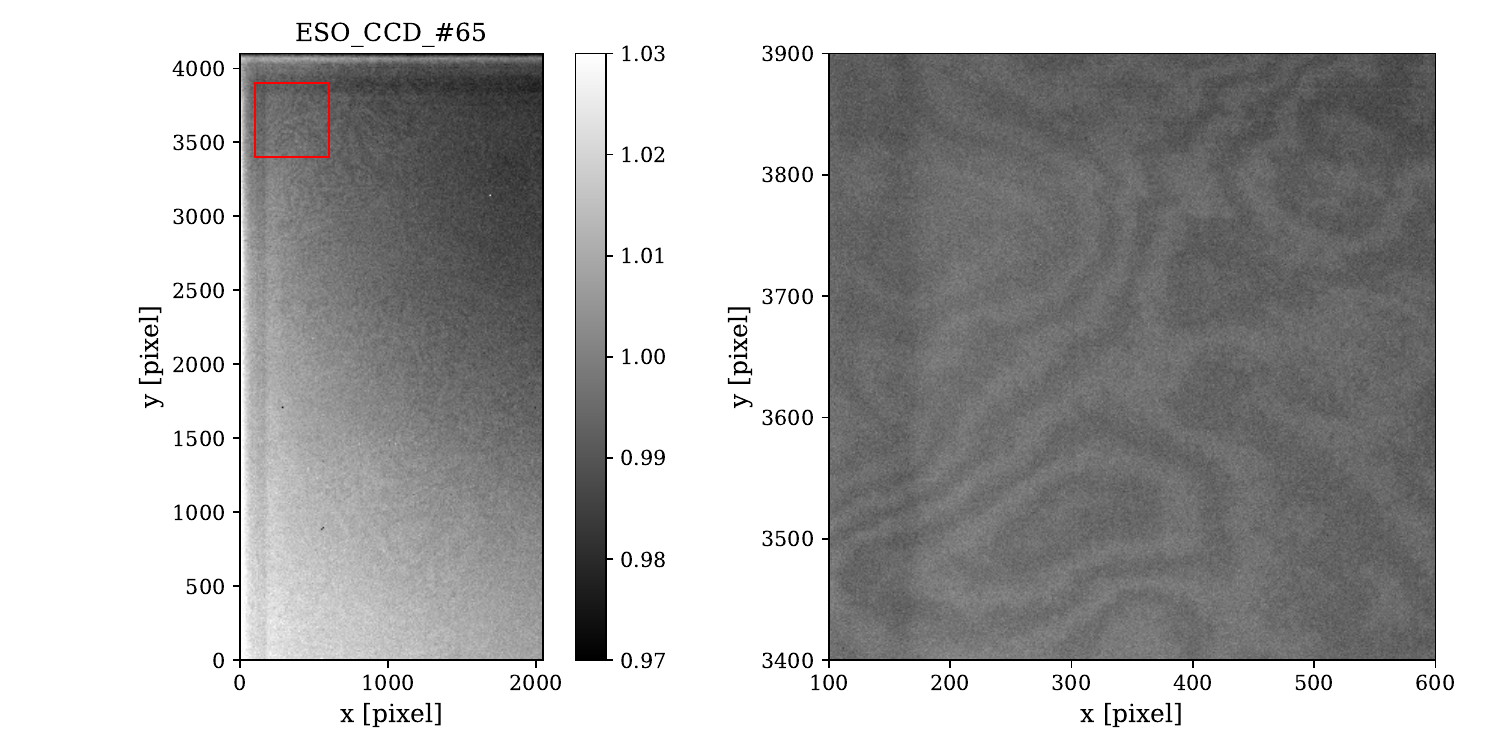}
    \caption{\modtexttwo{Flatfield (delta flat)} in the $r$-band obtained by averaging the science observations that were already flatfielded using the standard \textsc{AstroWISE} pipeline. The values shown in the colorbars represent the (dimensionless) relative change with respect to the original flatfield. The top panel shows the full mosaic. The bottom figures show the bottom-left chip with a zoom-in of the top-left corner of that chip (red square). Some structure is visible, likely due to variations in the illumination, as well as some low-level fringing.
 \label{fig:flat_r}}
\end{figure}

The resulting \modtexttwo{delta flat} for the $r$-band is shown in Fig.~\ref{fig:flat_r}. The top image shows the full mosaic. We observe a clear radial pattern, albeit with a small variation, which is caused by the illumination correction that is applied in the \textsc{AstroWISE} pipeline. The jumps between the chips arise because we normalise the individual chips to have a mean of unity. Unlike lensing studies that use these data to determine photometric redshifts, ensuring a consistent zero-point across the field-of-view is not essential for our aim: our objective is a smooth sky on average. To achieve this, we apply a zero-point correction to the background-subtracted images, which is discussed in Sec.~\ref{sec:ill_corr}. 

The bottom row in Fig.~\ref{fig:flat_r} shows the bottom-left chip, as well as a zoom-in of the top-left corner (red square). Although we started with already flat-fielded images, some structure is visible. This is likely caused by the variations in the illumination, as well as some low-level fringing. Moreover, after we apply this additional flatfield to the data we find that the background also shows features. In particular, gradients in the background persist. The data we used in this paper were obtained early in the survey, as fields overlapping with GAMA were prioritized. At that time, the baffling of the telescope was not optimal (this was corrected later). This is the likely cause for the remaining variation, caused by changes in the illumination. We decided not to attempt further improvements and accepted that this will limit our IGL measurements in the end. To make the images more homogeneous for the masking step, we subtract a constant background from each chip using the median of the pixels that are unmasked in the \textsc{SExtractor} segmentation image. \modtext{ This was done per frame (exposure) for the individual CCDs.}

Although the `reduced science frames' contained an initial astrometric solution, it needed to be refined (in the standard KiDS pipeline, they were astrometrically calibrated at this stage as well). We used \textsc{Scamp} \citep{bertin2006} using the Gaia DR2 \citep{brown2018} as reference. To map the distortion of the camera we used a second-order polynomial because the overall distortion of the camera was found to be small \modtexttwo{similar to the approach taken in \citet{Kuijken2019}}. We found that with this setup the residuals in the astrometric solution are negligible (about $0\farcs 01$ dispersion). We use \textsc{Swarp} without background subtraction to map the individual chips to a single image that is used for the measurements of the surface brightness profiles.
\modtexttwo{\subsubsection{Masking all sources}}
\label{sec:flat_mask}

Internal reflections result in ghost haloes \modtext{(for details on this feature and how they were masked, see \citealt{dejong2015}, and their fig.~4)}. These are very apparent near bright stars, but are in fact always present.
We chose to mask the reflection ghosts for very bright stars, for which the excess flux is clearly visible. This ensures that the most significant contributions are removed, while the remaining ghosts increase the uncertainty in our measurements somewhat. We determined the locations of the reflection ghosts as a function of position in the focal plane. We used the {\it Gaia} third Early Data Release (EDR3; \citealt{brown2021}) to estimate the fluxes of bright stars in the images, and masked the affected regions if the predicted magnitude  is brighter than $m=10.5$ mag in the filter of interest. Although the ghosts are shaped like a doughnut, we also masked the inner regions. We masked bad columns, and masked all stars brighter than $m_\mathrm{G}=16.5$ mag using the {\it Gaia} photometry with an aperture of radius $r_{\rm ap}=175-20\times (m-10)$ pixels, which ensures that most of the starlight is masked. 

Finally, some of the images suffer from erratic gain variations caused by a problem with one of the video boards\footnote{\href{http://www.eso.org/observing/dfo/quality/OMEGACAM/qc/problems.html}{http://www.eso.org/observing/dfo/quality/OMEGACAM/qc/problem s.html}}. We identified those images and masked these from our analysis \modtext{(these are referred to as `bad pixels' later).} 

The contributions of remaining objects, stars, and galaxies were masked using the \textsc{SExtractor} segmentation images. \modtext{ We use the same \sextractor run that was used to make the sky flat. To ensure masking the faint outskirts of the sources, we extend the marked regions in the segmentation image in this step. We explore the best setting in Sec.~\ref{sec:mask_test}.} The resulting masked images should only contain background, but occasionally objects are missed by \textsc{SExtractor}. These are readily removed by masking pixels with absolute values $>50$ counts (we adopt a zero-point where 1 count corresponds to $m=30$). When measuring the surface brightness profiles around BGGs, we unmask the pixels that correspond to the segmentation image of the galaxy of interest. 

\begin{figure*}
\sidecaption
\leavevmode \hbox{%
  \includegraphics[width=12cm]{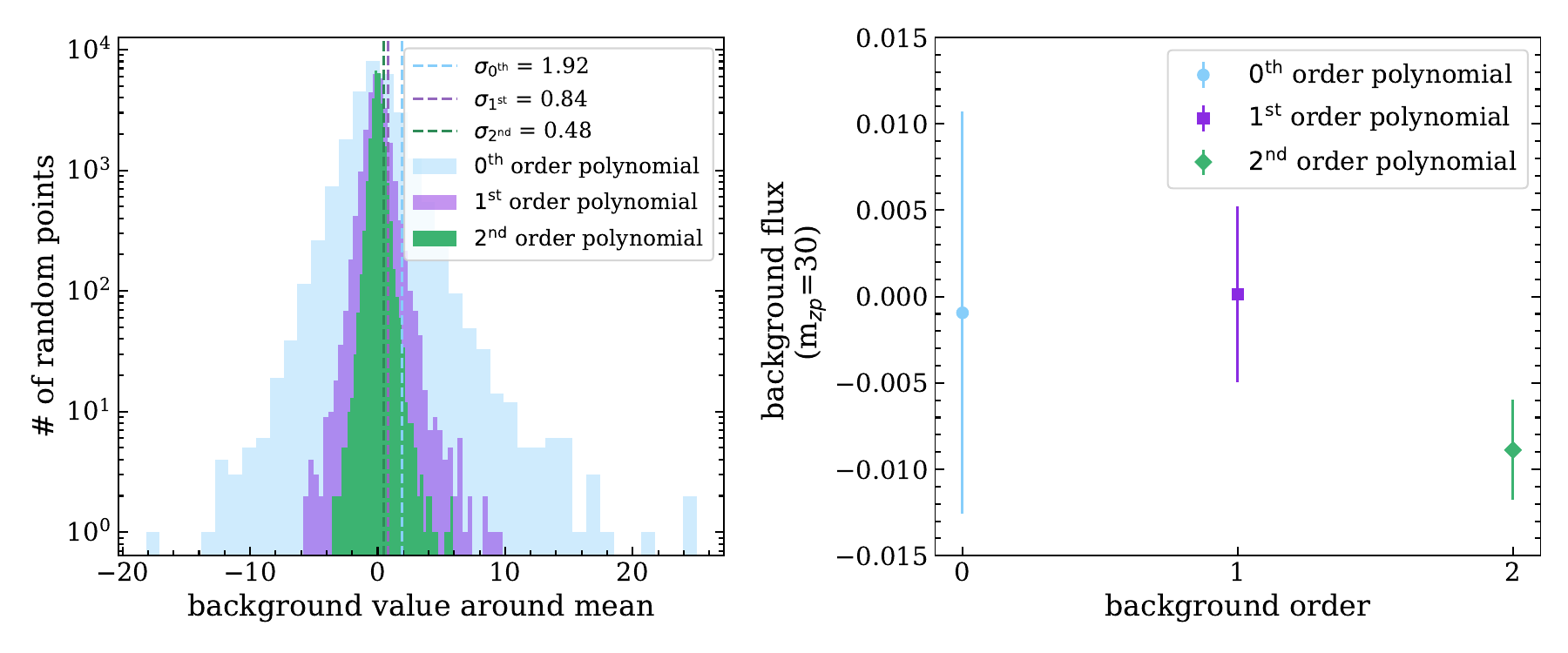}}
\caption{Left: histogram of the background values in randomly placed annuli on the field images with polynomial sky subtractions of order 0 (blue), 1 (purple), and 2 (green). The vertical lines denote the corresponding 1~$\sigma$ values, as shown in the upper right corner. The scatter to the background values is smaller for higher-order polynomial estimation of the background. It is clear that higher-order polynomials remove spatial variation in the background more efficiently. Right: mean and scatter of the mean background values for the different polynomial background estimations (indicated in the upper right corner). This again demonstrates how the scatter is gradually reduced for higher-order polynomial estimation to the background values.  
 \label{fig:bkg_m012}}
\end{figure*}

\subsubsection{Zero-point Correction}
\label{sec:ill_corr}

The non-uniform illumination due to the presence of additional stray light in the combined field images was corrected for all the photometric bands \modtexttwo{using the chip-by-chip background subtraction described in Sec.~\ref{sec:pipeline} (for more details on the stray light issue, see \citealt{dejong2017} and \citealt{Kuijken2019})}. However, this created a non-uniform zero-point (ZP) shift across each field image, which was corrected at this stage. For this, we selected \textsc{SExtractor} magnitudes (AUTO\_MAG) of the stars with $m \geq 16$ in our field images and measured the residual systematic magnitude differences compared to the KiDS DR4 source catalogue \citep{Kuijken2019}, which was corrected for all these systematics. The spatial variations of these differences were then fitted with a second-order, two-dimensional polynomial for a subset of the field images. \modtext{The distribution of the fitted polynomial coefficients for the non-uniform ZP-variation was consistent with less than 0.5 per cent variation for all the images we tested. We took the average fitted coefficients for the two-dimensional polynomial and created a `correction' image with the same pixel size of the field images.} Each of the field images was then divided by the correction image on a pixel level to obtain the ZP-variation-corrected images. As the background was already subtracted beforehand, this division does not affect the overall background level of the image, but makes the zero point of the sources spatially uniform. Using a second-order polynomial may leave some small-scale features in the photometric calibration, but this effect is mostly suppressed after stacking.

The absolute zero-point shifts of all the illumination-corrected images were finally measured by comparing the magnitudes of the stars in each field to the KiDS DR4 source catalogue. These shifts were accounted for while converting counts to surface brightness of the CG radial profiles \modtext{using eqn.~\ref{eq:zp_corr}.}

\subsubsection{Exploring the bias from the sky subtraction}
%
\label{sec:bgtest}

The masked images have a low background (we subtracted a constant value from each chip before combining them using \textsc{Swarp}), but now that most sources are masked, we need to improve the sky subtraction as there are still remaining gradients in the background. In this section, we explore the impact of subtracting a low-order polynomial from each chip. 

To quantify the performance, we measured the scatter in the background estimates in randomly placed annuli with inner and outer radii of 100 and 150 arcseconds, respectively. We measured the mean and scatter in the values. Especially the latter is of interest, as a lower scatter implies that we can measure the surface brightness profiles to larger radii. However, even though the images have been masked rather aggressively, the main concern is that the diffuse light around the CG may still impact the estimate. This will be more relevant for higher-order polynomial fits to the background. We therefore explore the impact on the galaxy profiles in Sect~\ref{sec:bkg_on_galaxy_profile} as well because, in that case, the data are weighted differently. Nonetheless, focusing on the background estimates alone will provide a first indication of the performance of our pipeline.

Figure \ref{fig:bkg_m012} shows the scatter of the background values in the randomly placed annuli described above for three different polynomial background estimations: zeroth order (blue), first order (purple), and second order (green) in the left panel. The right panel shows the mean values of the background with the standard error to the mean for each of the polynomial background estimations as indicated in the top right corner. For each of the background estimations, 30 random annuli were placed in every field image and exposure. As all the \textsc{SExtractor}-detected sources, bad pixels containing too high or low values, and bright foreground stars were masked in the image before placing the random annuli (as described in sec. \ref{sec:pipeline}), some of the annuli had a large fraction of the possible pixels masked. Especially if the location of the annulus was near any of the bright stars, up to 90 per cent of the possible pixels in the annulus were masked in some cases. This gave rise to a higher noise and scatter to the mean background value for such annuli. To avoid such cases, we selected only those random annuli where at least 40 per cent of the total possible pixels in the annulus were unmasked for the background value estimation. 

As is clearly visible from Fig. \ref{fig:bkg_m012} (values of $\sigma$ specified in the legend of the left panel and the error bars in the right panel), there is about a factor of two improvement in the scatter of the background values for each increased order of polynomial to subtract the background. This means that the precision of the background value estimation increases with increased order of the polynomial as structures in the background are removed, and for the cases we considered, a second-order polynomial results in the lowest scatter to the estimated value. However, the mean value of the background is the most biased for the second-order polynomial out of the three cases (value is less than 0 at $\sim3\sigma$ distance from the mean). \modtext{To select the most efficient background estimation out of these three options for our purpose, we also need to consider the impact of the sky background estimate on the extended galaxy profiles.}
We explore this further in Sec.~\ref{sec:bkg_on_galaxy_profile}.

\section{Light profiles}
\label{sec:galaxy_profile_and_masking_sat}

\subsection{Impact of background subtraction on extended galaxy profiles}
\label{sec:bkg_on_galaxy_profile}

An over-subtracted sky background will particularly affect the outer edges of galaxy light profiles \citep[e.g.][]{chamba2022}. If a model over-subtracts the background light, then the faint and diffuse light at the outer edge of a galaxy will be removed as background light. As a result, a galaxy's surface brightness (SB) profile will reach \modtext{zero at a closer radial distance from the  centre of the galaxy.} \st{and the profile will have more negative values towards the outskirts}. The background subtraction that preserves the most flux in the outermost radii of an extended source will therefore be the most accurate estimation of the global sky background. To test the accuracy of the background subtraction models, we checked the extended profiles of bright sources with each of the background estimation cases ($0^{\mathrm{th}}$, $1^\mathrm{st}$, and $2^\mathrm{nd}$ order polynomials).
\begin{figure}
\centering
\leavevmode \hbox{%
  \includegraphics[width=0.9\columnwidth]{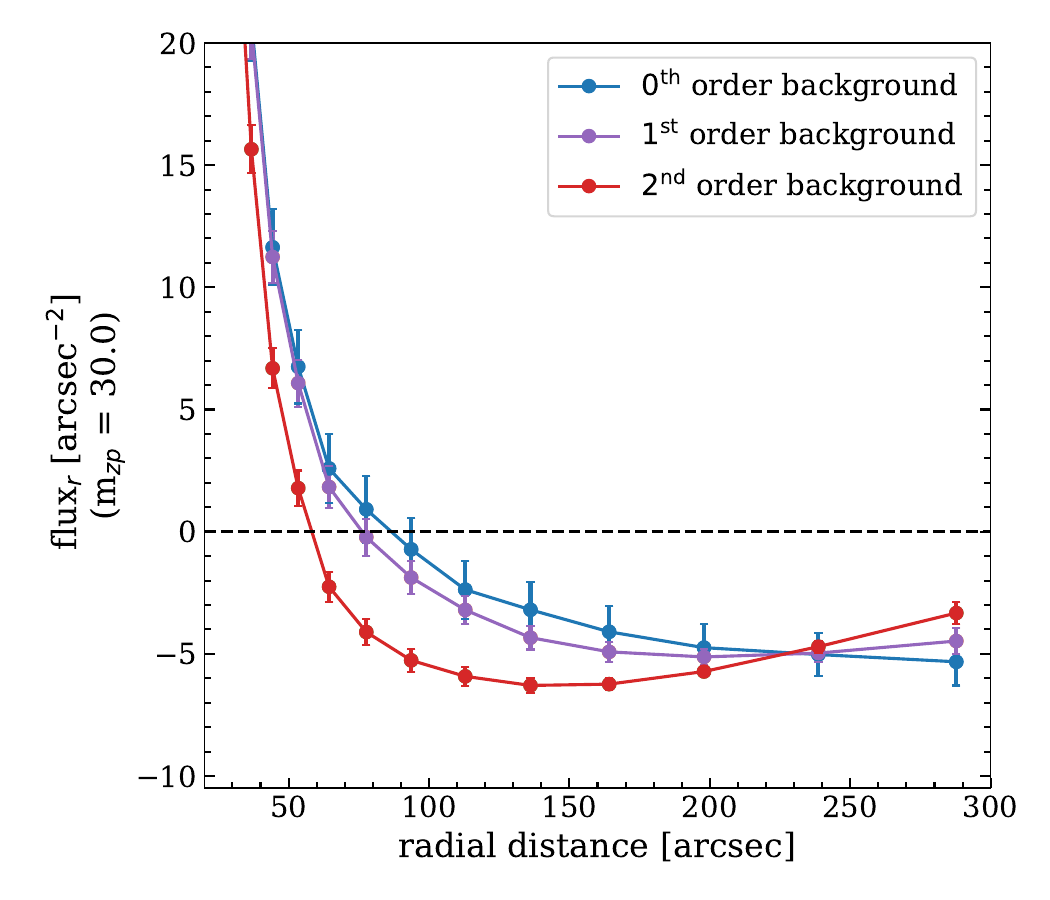}}
\caption{Surface brightness profiles of the group central galaxies beyond 20 arcseconds radial distance from the centre for 0$^{\mathrm{th}}$ (blue), 1$^{\mathrm{st}}$ (purple), and 2$^{\mathrm{nd}}$ (red) order polynomial background estimations, respectively. In all the profiles, error bars indicate $1\sigma$ uncertainties on the mean. All three profiles are the same within 20 arcseconds (not shown). Beyond that, however, higher-order background estimations over-subtract the background compared to the lower-order ones. This is the most prominent for the 2$^{\mathrm{nd}}$-order polynomial estimation of the background. 
 \label{fig:bkg_p012_on_sb}}
\end{figure}

For this test, we selected GAMA groups at $0.09<z<0.15$ with a bright central galaxy ($M_{\mathrm r}\leq -23$ mag) and constructed their stacked SB profiles for each of our background models. The extended part of the resulting SB profiles is shown in Fig.~\ref{fig:bkg_p012_on_sb}. As is visible from the figure, the $0^{\mathrm{th}}$ order polynomial fit to the background retains the most light at the outer edge of the galaxies, with higher order polynomial fits retaining consecutively less light, and $2^{\mathrm{th}}$ order fit \modtext{having the least amount of light retention at the outskirts}. Considering the tests demonstrated here and in Fig.~\ref{fig:bkg_m012}, the $1^\mathrm{st}$ order background subtraction seems to have a reasonable performance in both cases. However, none of the background subtraction models is unambiguously preferred above the rest. In our following analyses and tests, we therefore use all three background-subtracted images and compare their performances. 

\subsection{Masking the satellites}
\label{sec:mask_test}

Along with the central galaxies, large satellite galaxies in galaxy groups and clusters can also have extended light, albeit a smaller amount. In a stacking analysis, such residual satellite light can result in a systematically higher diffuse light estimation than the actual amount. While measuring the IGL, it is therefore essential to ensure that light from satellite galaxies is completely masked out. We obtain the initial masking to the satellite galaxies and other sources in the image from the segmentation map of the \sextractor output. \modtext{However, \sextractor can fail} to include the fainter light distribution around the sources, which is more visible for the satellite galaxies and projected nearby galaxies along the line of sight. To address this, we extended the source masks obtained from the segmentation map. An increased mask size is expected to cover possible faint light surrounding the sources, but it also has a potential risk of masking the faint IGL signal and reducing the total light fraction in the IGL. We explored different levels of mask extension to identify an optimum extension for our analysis. We found that masks from the original segmentation map and less than 10-pixel extensions (4 or 6 pixels) are too small to exclude residual extended light from satellites and retain small-scale irregularities in the extended light. Likewise, a larger (20 or 40 pixels) extension of the segmentation map over-subtracts the extended light. Considering both issues, we concluded that a mask extension of at least 10~pixels is needed to lower the contribution of extended light from satellites in the CG+IGL content. We therefore used this 10-pixel mask extension throughout this work where any masking was used.

\subsection{Point spread function}
\label{sec:psf}
\modtext{The PSF of updated KiDS images was constructed following a similar method as \citet{Montes2021}, \citet{infante-sainz2020}, and \citet{Zhang2019}}: we constructed the PSF by stitching profiles of bright and faint stars in different sections because very bright stars are saturated in the central region and light from the fainter stars at larger radii are not detectable with a high signal-to-noise ratio. 

We started by running \sextractor \citep{bertin_arnouts1996} on each of our updated KiDS field images to obtain the source catalogues. To determine which stars are suitable for which parts of the PSF we used the half-light radius (`FLUX\_RADIUS') and the magnitude (`MAG\_AUTO') parameters of the detected sources. We also used the stellarity index (`CLASS\_STAR') provided by \sextractor to classify if an object is a star (1) or a galaxy (0). To select the unsaturated stars, we used all objects with CLASS STAR larger than 0.65. \modtext{ We verified that this CLASS\_STAR value maximized the selection of star-like objects in the MAG\_AUTO vs FLUX\_RADIUS parameter space. All of the unsaturated stars lie in a narrow FLUX\_RADIUS range with slightly varying fluxes because of the PSF FWHM. Our chosen CLASS\_STAR could separate these from the extended sources at each flux level. 
The saturated stars were chosen only from their magnitude and size. Further details on the star selection are given in the following sections.}

\subsubsection{Estimating core, intermediate, and outer parts}

We divided the PSF into four sections: core, intermediate, outer 1, and outer 2. The half-light radius and the aperture magnitude were used to select which stars construct which part of the PSF. The core section is constructed from the SB profile of bright unsaturated stars between magnitudes 16.5 and 18 mag. Slightly brighter stars with saturated central regions but extended profiles out to larger radii were used for the intermediate part of the PSF. These stars for the intermediate profile had magnitudes between 14 and 15.5 mag. \modtext{The outer 1 and outer 2 sections consist of the brightest stars in the fields of view, with magnitudes between 12 and 14 mag. We chose to divide the brightest stars into two samples based on their median FLUX\_RADIUS to make a better transition from the outer 2 part to the intermediate part. In the combined PSF shown in Fig.~\ref{fig:psf_global} (from 192 field images $\times 5$ exposures), there are $\sim 5000$, $\sim 1000$, $\sim 400$, and $\sim 200$ stars in the `core', `intermediate', `outer 1',  and `outer 2' sections, respectively. }


\subsubsection{Stacking and stitching different parts of the PSF}
\label{sec:measure_psf}
\modtext{To estimate the PSF in each of the four sections (core, intermediate, outer 1, outer 2), we made cutout stamps of stars, stacked the stamps, and calculated their radial SB profiles from the stacked images. The sizes of these stamps vary based on the part of the PSF the stars were used for. For the core part, we did not need a profile that extends to the full range of the PSF, and made $100\times100$ pixel ($21.4\arcsec\times 21.4\arcsec$) stamps centred on the stars. For the intermediate stars, we made $500\times500$ pixel ($1.8\arcmin\times 1.8\arcmin$) stamps; and for the stars in the outer~1 and outer~2 regions, we made $1000\times1000$ ($3.6\arcmin\times 3.6\arcmin$) pixel and $2000\times2000$ pixel ($7.2\arcmin\times 7.2\arcmin$) stamps, respectively.}

We used the segmentation map from \sextractor to exclude all other sources except for the central star in each stamp. To exclude light that is not masked by the segmentation map, we also applied a $3\sigma$ clipping method \modtext{(excluding any pixel that has a value above or below $3\sigma$ from the median pixel value in the masked stamp, where the masks also included the central stars). We also excluded all the stars from the sample that have a brighter star in the stamp. Another influence on the background of the KiDS field images is the large reflection ghost caused by large saturated stars.} These ghosts cause the surface brightness to be elevated in certain parts of the PSF. This effect was tested, and all the fields with such ghosts were removed from the stack in the final PSF estimation.

\begin{figure}
\centering
\leavevmode \hbox{%
  \includegraphics[width=0.9\columnwidth]{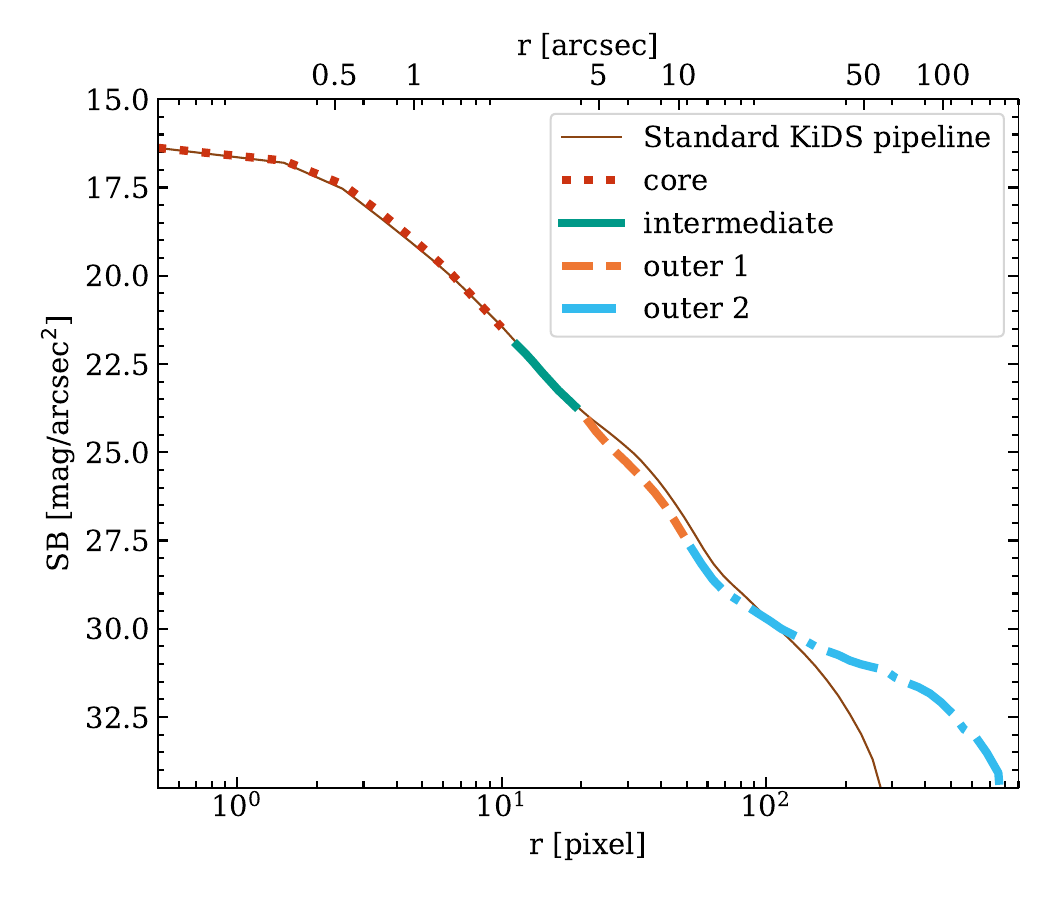}}
\caption{Stitched point spread function (PSF) from combining stars within different magnitude ranges (see text) from all fields with updated background-subtracted images. The colours and line styles indicate the radial range where different groups of stars (core, intermediate, outer 1, and outer 2) contributed to constructing the total PSF. The brown solid line shows the PSF constructed in the same way as above from standard KiDS data-release 4 images. The excess of faint light beyond 200~pixels in the PSF from the updated pipeline indicates the missing light in the standard KiDS pipeline.
\label{fig:psf_global}}
\end{figure}

The selected stamps were stacked, and the radial SB profiles of the stacked images were measured. \modtext{Finally, the four partial profiles were stitched together for each pointing and exposure. The stitching was done by selecting the part of the profile before it drops discontinuously from the continuum in one section (e.g., core) and replacing it with the next section (e.g., intermediate after the core) from there. A common area of 10 (inner) to 20 (outer) pixels of overlap between two consecutive sections was maintained during the stitching to ensure continuity of the profile. After the stitched PSFs were calculated for each pointing and exposure, they were stacked to obtain the final PSF. We only considered the 1D PSF in all measurements. The final PSF profile is shown in Fig.~\ref{fig:psf_global} along with the PSF from the original KiDS DR4 field images. Comparing these two PSF profiles, it is clear that the updated background-subtracted images can detect PSF flux at larger distances compared to the images from the standard pipeline. A slight mismatch at $\sim 30$~kpc is also visible upon closer inspection. The KiDS PSF is measured by averaging the PSF of 1004 pointings $\times$ 5 exposures. The modified background PSF is measured from averaging 192 $\times$ 5 exposures (the pointings containing the KiDS+GAMA group sample). The larger sample also has a different distribution of seeing compared to the smaller one. The slight mismatch in the PSF shown at $\sim 30$~kpc in Fig.~\ref{fig:psf_global} likely comes from the effect of the difference in sample size and their seeing distribution. Moreover, the scale on which the PSF profile differs from the standard KiDS pipeline roughly corresponds to the mesh size used by SExtractor, which may also be relevant for the mismatch. } 

\section{Prediction from simulations}
\label{sec:sim_prediction}

In \citet{ahad2023}, we prepared mock observations from the Hydrangea simulations \citep{bahe2017hydrangea,barnes2017cluster}, a suite of cosmological hydrodynamic zoom-in simulations of $24$ massive galaxy clusters with virial mass between $10^{14.0}$ and $10^{15.4}~\msun$ at $z = 0$. Each of the zoom-in regions includes the large-scale surroundings of the clusters to $\geq 10$ virial radii ($r_{200c}$) at $z=0$, containing many group-mass haloes in addition to the central clusters. The simulations were run using the AGNdT9 calibration of the EAGLE galaxy formation and evolution code \citep{schaye2014eagle}. Different subgrid physics models were used to simulate astrophysical processes that originate below the resolution scale of the simulation, including star formation, star formation feedback, radiative cooling and heating, stellar evolution, black hole seeding, growth, and feedback. For details about the simulation model, hydrodynamics scheme, and comparison of the model to observed galaxy properties, see \citet{schaye2014eagle, schaller2015eagle, crain2015eagle,bahe2017hydrangea} and references therein.

The group sample in \citet{ahad2023} was chosen to be comparable to our baseline KiDS+GAMA group sample with $0.09\leq z\leq 0.15$. The $u-$ and $r-$band mock observations were also made with comparable noise levels to the KiDS data. \modtext{However, to make a better comparison to the KiDS+GAMA group analysis, we still need to account for the smearing by the point spread function (PSF) of the KiDS images \modtexttwo{(with the updated pipeline used in this work)} on the stacked SB profiles from Hydrangea groups.} 


The PSF of the instrument distributes the bright light at the core of stars and galaxies to the outer region. As a result, the diffuse light at the outskirts of bright group CGs may get excess contribution from the smearing of the central light. We tested the effect of the PSF on the IGL measurement in two ways. First, we measured whether the fraction of light in the IGL changes due to the PSF. Second, we tested whether the radial range of IGL detection is affected by the PSF.

\begin{figure}
\centering
\leavevmode \hbox{%
    \includegraphics[width=0.9\columnwidth]{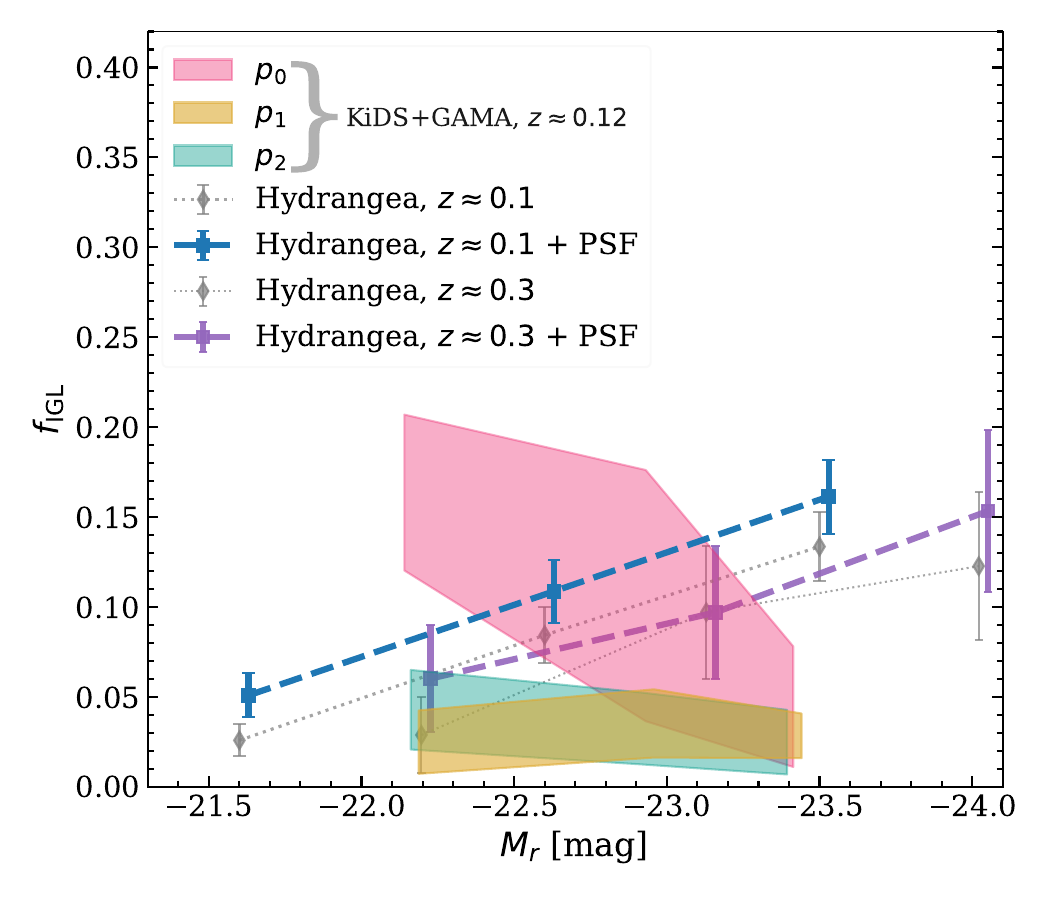}}
    \caption{Fraction of light in IGL compared to the total group light ($f_{\mathrm{IGL}}$) of our KiDS+GAMA group sample at $0.09<z<0.15$ in the narrow bins of central group galaxy \modtexttwo{(CG) magnitude in $r-$band ($\magr$)}  they were stacked. The values along the x-axis indicate the mean $\magr$ of the corresponding bins. Pink, yellow, and cyan shaded regions show the upper and lower limit of $f_{\mathrm{IGL}}$ for the $0^\mathrm{th}$ ($p_0$), $1^\mathrm{st}$ ($p_1$), and $2^\mathrm{nd}$ ($p_2$) order background-subtracted images, respectively. Here, the $p_0$ measurement is the upper limit of $f_{\mathrm{IGL}}$ for each magnitude bin at this redshift bin, while $p_1$ and $p_2$ measurements provide a lower limit. Details on how we define the upper and lower limits for each of the measurements are discussed in the text. \modtext{The dotted and dashed lines show the effect of the PSF on the measurements at redshifts 0.1 and 0.3 from the Hydrangea simulations.}
    \label{fig:f_igl_all_p_z0915}}
\end{figure}

To measure the effect of the PSF on the IGL fraction measurement, we took stacked SB profiles of Hydrangea groups at relevant bins of their $r-$band magnitudes from \citet{ahad2023} and fitted a single de Vaucouleurs profile to the CG to separate the IGL. The fitting was done for both unconvolved and PSF-convolved SB profiles. The fraction of light in IGL ($f_{\mathrm{IGL}}$) from the PSF-convolved profiles were measured using the corresponding PSF-convolved total group light profiles. \modtext{In both cases (CG+IGL and total group light profiles),} the stitched PSF profile (from Sec.~\ref{sec:measure_psf} and Fig.~\ref{fig:psf_global}) was normalised by the total flux in the PSF before the convolution. The $f_{\mathrm{IGL}}$ measurements are shown in Fig.~\ref{fig:f_igl_all_p_z0915} for both convolved (blue, purple) and unconvolved (grey) cases, with errorbars showing \modtext{the standard error of the measurements given the sample size}. As is visible from Fig.~\ref{fig:f_igl_all_p_z0915}, the values do not change much considering the error bars. However, for groups with a brighter CG, $f_{\mathrm{IGL}}$ seems to be slightly underestimated due to the effect of the PSF at $z\sim0.1$.  

\modtexttwo{The measurement of $f_{\mathrm{IGL}}$ (CG+IGL and total group light both) was done until an SB limit of 30 mag/arcsec$^2$. This limit was chosen to match the SB limit of stacked radial profiles of the KiDS+GAMA groups, which was chosen to have at least signal-to-noise ratio of 5.} Along with measuring $f_{\mathrm{IGL}}$, we also checked how the radial range of IGL is affected by the PSF convolution within the SB limit. For this, we measured the radial distance from the group centre within which 90 per cent of the IGL is enclosed ($r_{90}$). For the three magnitude bins we considered, $r_{90}$ increased with CG magnitude. Compared to the unconvolved profiles, values for $r_{90}$ were about 100~kpc larger in the PSF-convolved profiles. For the convolved profiles, $r_{90}$ for the three considered magnitude bins were about 260, 380, and 470~kpc, respectively. 

The key conclusion from this test is that the PSF does increase the radial IGL detection range. However, it does not increase the measured $f_{\mathrm{IGL}}$ significantly because the same PSF effect is also present in the total group light. As a result, the impact of the PSF is not large for the CG magnitude range we consider, but we consider this effect in our comparison nevertheless.

\section{The intragroup light in GAMA+KiDS groups}
\label{sec:igl_chapt}

To quantify the intragroup light (IGL) in our group sample, we first measured the radial surface brightness (SB) profiles of each group from the reprocessed KiDS field images. A representative sample of the profiles were then checked visually to flag any profile that could potentially introduce bias in our measurements. Only the unflagged profiles from each redshift range were then grouped in bins of central galaxy (CG) magnitude before stacking and measuring the IGL in stacked profiles. Details of this process are given below.

\subsection{Radial surface brightness profiles}
\label{sec:sb_prof_prep}

For each of the KiDS pointings, we have five exposures in $g, r,$ and $i-$bands and four exposures in $u-$band. Our results are based on the $r-$ band data because these are the deepest among the four bands. We analysed each exposure separately. For each group, we created a $2\times2$~Mpc cutout (at the appropriate redshift) centred on the group CG. \modtext{The group CG location was taken from the GAMA \texttt{G$^3$CFoFv08} catalogue (more details on the catalogue are provided in Sec.~\ref{sec:gama_data_intro}). As the GAMA group catalogue was not based on the KiDS imaging we used in this work, there can be a small variation of the exact photometric centres of the CGs. The \sextractor output catalogues based on the KiDS imaging provide a more accurate central pixel for the group CGs. We compared the CG centre location from the GAMA catalogue and the \sextractor output catalogue, and updated the CG centres with the \sextractor-provided location in case there was a difference between the two. The $2\times2$~Mpc cutouts are centred on these updated centres.} 

\modtext{We applied the bright star and bad column masks (more details on these masks are given in Sec.~\ref{sec:pipeline}) on the cutouts, and measured two different SB profiles from each cutout: (i) the CG+IGL SB profile that had all the sources masked except for the CG; and (ii) the total group SB profile that had all the sources masked, except for the group member galaxies (including the CG). The masking procedure for satellites and non-member sources is explained in Sec.~\ref{sec:mask_test}. We created the azimuthally averaged radial SB profiles using circular apertures centred at the CG of each group to 1\,Mpc radial distance. The central parts of the profiles were linearly binned from one to 10 pixels for better sampling; logarithmic binning was used beyond that. The zero-point correction for the corresponding field and exposure (as discussed in Sec.~\ref{sec:ill_corr}) was accounted for during the SB profile measurement as follows: }
\modtext{\begin{equation}
    SB\ [mag] = -2.5* \log_{10}(SB\ flux) + 30 + \textrm{zp-correction}
    \label{eq:zp_corr}
\end{equation}}
\modtext{The standard error to the individual SB profile was measured using the formula $\sigma/\sqrt{n}$, where $\sigma$ is the standard deviation of the unmasked pixels in each radial bin, and $n$ is the number of unmasked pixels in each radial bin. }

During the total group light measurement, we applied an additional distance selection for the satellite galaxies considered for the total group light. This selection was made to account for the uncertainty in the group member assignment that comes from the aggregation of low-mass groups in the FoF halo finder algorithm. By comparing the KiDS+GAMA groups to the BAHAMAS simulations, \citet{Jakobs2018} found that aggregation of multiple low-mass groups into one was present in 37 per cent of groups/clusters in their sample. To limit the inclusion of such potentially wrongly included satellite galaxies, we use the $\mathrm{Rad}_{50}$ and $\mathrm{Rad}_{100}$ parameters of the GAMA group catalogue, which indicate the distance from the group CG within which 50 and 100 per cent of the group members are located. Looking at the distribution of the $\mathrm{Rad}_{100}/\mathrm{Rad}_{50}$ ratio, we found that in about 20 per cent of our group sample, there is at least one satellite galaxy that is more than 3$\times \mathrm{Rad}_{50}$ away from the CG. These secluded distant satellites are highly likely to be wrongly assigned group members due to the effect of aggregation and including these galaxies in the total group light can potentially bias the IGL fraction measurements to a lower value than its actual amount. Therefore, we only considered satellites within 3$\times \mathrm{Rad}_{50}$ from the group CG in our total group light profiles. 

\subsection{Profile selection to lower measurement bias in stacking}
\label{sec:good_prof}
During the stacking analysis, significant outliers can bias the overall measurement. Therefore, to ensure the robustness of our measurement, 25 per cent of the SB profiles\footnote{\modtext{There were $\sim$500 groups with 5 exposures each in each redshift bin. The visual inspection was done on a random sample of 25 per cent of 500$\times$5 individual exposures. As the profiles were computed for each of the five exposures for each group separately, a randomly selected sample of 20 per cent of the profiles is likely to represent at least one of the exposures for each group. We selected 25 per cent to increase the chance that every group was checked at least in one of the exposures.}} (randomly selected from the complete sample) were visually inspected to identify possible causes for having an outlier, and a `flag' value was given to each of the SB profiles (of the complete sample). 

\modtext{A positive integer value of the flag was assigned to every profile based on the type of irregularity in the SB profiles. All the profiles were flagged based on the criteria that were defined by inspecting 25 per cent of the profiles.} A flag value of 0 indicates no issue and a good profile. The most frequent reason to flag was the fraction of masked pixels in a cutout. If a group cutout or its central 25 per cent area had at least half of the total pixels in that area masked (bad pixels, stars, or other galaxies), it was flagged. Masked fractions in both the cutout and its central area were flagged quantitatively. If the masked fraction was low in a group, but the mask overlapped with part or all of the CG and resulted in a non-existent segmentation map at its CG location, \modtexttwo{it was also flagged}. If a CG SB profile had its brightest point shifted from the centre or had a significantly low central flux count ($\leq 200$, compared to more than $\sim1000$ for a standard profile), it was also flagged. These particular cases happened mostly due to a partially masked CG, which was checked and confirmed for all the group cutouts that were assigned with the corresponding flag value. All of these flagging reasons are connected with our conservative masking procedure. In addition, SB profiles that had empty values (defined as `not a number' or `NaN', mainly in cutouts where masked areas covered a ring-like pattern around the CG for the presence of many bright sources around it) for multiple radial distances and CG+IGL SB profiles that had high scatter (larger than the median variation of flux count beyond 400~kpc for each considered group sample, usually $\sim1.0$ flux count variation) in the far outskirts were also flagged. Finally, if any CG+IGL SB profile was flagged in $\geq3$ of the five available exposures, the rest were flagged for lack of reliability. \modtext{After removing the flagged profiles, the good profiles were all checked visually to confirm that there were no strong outliers.} 

Another selection criterion we applied was removing GAMA groups that potentially have an ambiguous CG in the GAMA catalogue. In \citet{ahad2023}, we demonstrated that if the group CGs were selected based on the galaxy halo mass\footnote{The galaxy halo mass here is different from the halo mass of the group which the galaxy is a member of. The galaxy halo mass was computed from the galaxy stellar mass and the (galaxy-colour-dependent) stellar-to-halo-mass-relation given by eqn.~7 of \citet{Bilicki2021}.} instead of selecting the brightest galaxy at the centre of light distribution, about 20 per cent of the GAMA groups in our sample would be assigned a different CG (predominantly a red one instead of a blue one). We also showed in \citet{ahad2023}, based on our mock observations from Hydrangea simulations, that such miscentring can slightly suppress the IGL measurements. However, the small suppression is inferred from a simulated sample analysis, for which we had information about the `true' halo centre, which is not the case for observational data. Therefore, we chose only the GAMA groups in our sample that did not have a re-assignment of their CG based on the galaxy halo mass. This selection lowered our sample size by a further 10 per cent. \modtexttwo{Because of the large initial group sample, even after applying our strict selection criteria, we had 323, 393, and 296 ($\times5$ exposures) groups in the low to high redshift bins, respectively.}

\subsection{Sub-stacking based on BGG luminosity}
\label{sec:sb_mr_bins}

In \citet{ahad2023}, we found that, based on mock images of groups (at $z=0.1$ and with halo masses $12.0\leq \log_{10}[M_{200}/\msun] \leq 14.5$) in the Hydrangea simulations, the IGL content has a positive correlation with the luminosity of the group CG. We also found that while stacking multiple group CGs, binning them as a function of the absolute magnitude of the CGs preserved the underlying IGL fraction trend \citep[fig.~8 of][]{ahad2023}. We utilised this result in this work and stacked group CGs of similar absolute magnitudes. 

To keep a uniform range of absolute $r-$band magnitude ($M_{\mathrm{r}}$) in our different redshift samples, groups with CG $\magr$ between -21.5 and -23.5 mag were divided into three bins. The bin widths were selected to have a similar number of groups in each bin. \modtext{The SB profiles of the groups in each redshift and magnitude bin were first normalised to have the redshift-corrected flux values at the mean redshift of the sample}. Also, the radii in each bin were converted to physical kpc at their corresponding redshifts and rescaled to the physical kpc at the mean redshift of the sample. Finally, the profiles were stacked to obtain the mean profile in each redshift and magnitude bin for the CG+IGL and total group light. \modtexttwo{Since we take narrow bins in CG magnitude (which is correlated with the group halo masses, and therefore their virial radii), the stacked profiles along the rescaled physical radii at the same redshift have comparable virial radii. Therefore, profile shapes within the same redshift and CG bins are self-similar as a function of distance from the CG centers.}

\begin{figure}
\centering
\leavevmode \hbox{%
  \includegraphics[width=0.9\columnwidth]{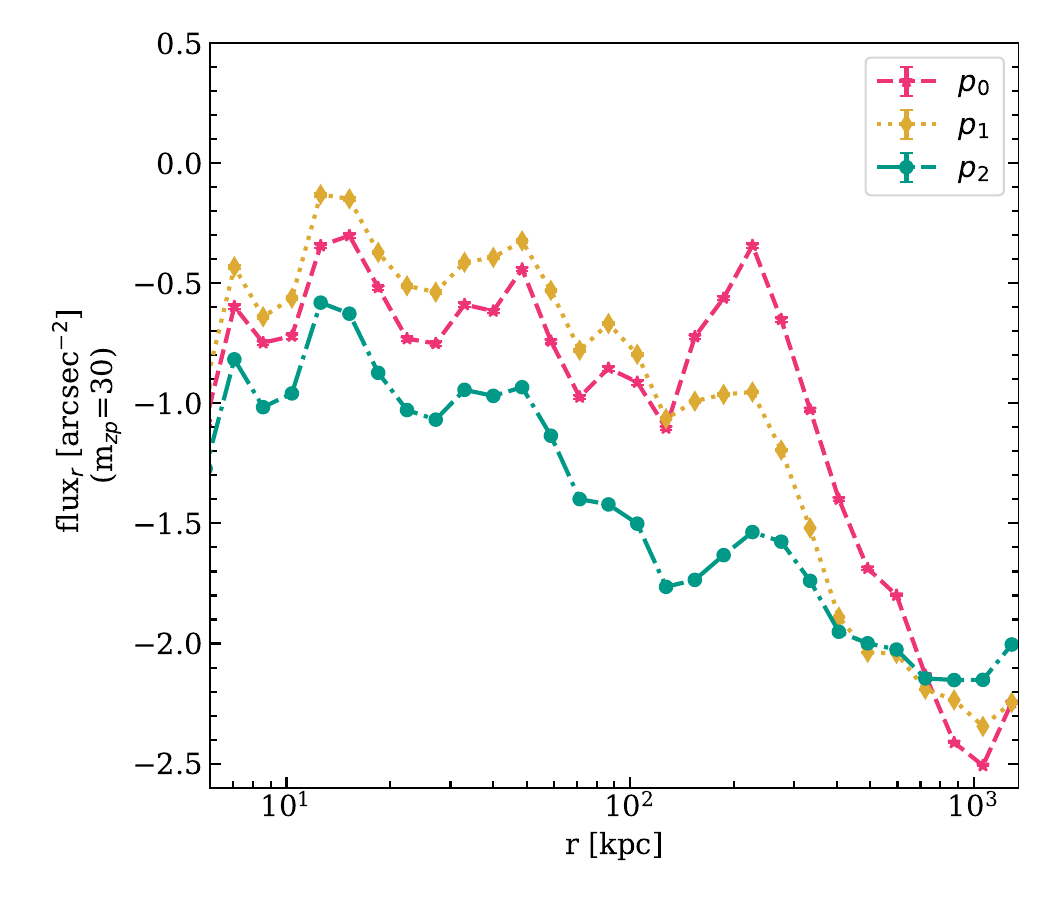}}
\caption{Surface brightness profile of the mean residual background at random points for the three background-subtracted field images used in this work. The error bars indicate the $1-\sigma$ scatter to the mean of the stacked background profiles. From the central region out to 100~kpc, the variation for $p_0$ and $p_1$ is similar, with $p_2$ having a radially increasing offset from the other two. Beyond 200~kpc, the $0^{\mathrm th}$ order background-subtracted images ($p_0$) have a steeper downturn than the other two. While $p_1$ has a similar trend as $p_0$, it is less steep. The downturn in $p_2$ is more gradual compared to the other two. However, for all three cases, the background reaches comparable values at $\sim$1~Mpc distance from the centre.
 \label{fig:res_bkg_p012_profiles}}
\end{figure}

\subsection{Residual background subtraction}
\label{sec:res_bkg_sub}

Instead of a local background-subtraction in the standard \textsc{AstroWISE} pipeline, our approach of a chip-by-chip global background subtraction minimizes any remaining flux pattern in the background (e.g. as shown in Fig.~\ref{fig:flat_r}). However, as Fig.~\ref{fig:flat_r} shows, there are some small-scale patterns at the chip edges that are difficult to remove even with our updated background subtraction model. Due to the presence of stray light \modtext{(e.g., from the reflected light of the Moon and planets)}, there are also some global remaining patterns in the joined field image. These patterns result in a non-flat background in the field image, which we verified to be comparable in all the field images considering the uncertainties. We accounted for this residual background by measuring a background SB profile and subtracting this from the stacked SB profiles of groups.

To do so, we measured the residual background profile at a random location for each of the group profiles. The random locations were obtained by taking the pixel location of a group CG in one pointing and measuring SB profiles at the same pixel location and cutout size in a different pointing. While preparing these profiles, all the \sextractor detected sources were masked with extended segmentation maps, and the masking was similar to how the group profiles were made. No pointing-and-exposure combination was used more than once to create the background profiles. Because we use the pixel location and cutout size of a group to create the background SB profiles, any global background pattern that may be included in the group SB profile due to its location (e.g., the centre of the image or edge of a chip), is accounted for in the residual background profile. Finally, the background profile fluxes were adjusted for the zero-point correction, and the radial range was adjusted to the appropriate physical kpc units before stacking and subtracting them from the group SB profiles. 

Figure~\ref{fig:res_bkg_p012_profiles} shows the mean residual background profiles in the field images for the three different background-subtraction models. Beyond 200~kpc, all the profiles shown have a downturn towards the end. For the $0^{\mathrm th}$ ($p_0$) and $1^{\mathrm st}$ ($p_1$) order background-subtracted images, the background value is stable with a small scatter to 200~kpc and then shows a sharp downturn out to 1~Mpc. For the $p_2$ background profile, however, there is an overall trend of lower values with increasing radial distance from the centre. After subtracting this residual background profile, \modtext{we reached an SB limit of 30~mag/arcsec$^2$ with SNR$\geq5$ for all the stacked group profiles.}

\subsection{Fraction of light in IGL}
\label{sec:f_igl_in_mr_bins}

The fraction of light in IGL compared to the total group light ($f_{\mathrm{IGL}}$) in each redshift and magnitude bin was measured in a similar process as done in \citet{ahad2023} using a single de Vaucouleurs (SD) profile fitting to keep the comparison consistent. We used the following steps:

\begin{enumerate}
    \item Any residual flat background light at the far outskirts (beyond 500~kpc from the group centre) was removed by fitting a constant background to the outer profile.
    \item The CG profile was fitted using a single de Vaucouleurs (SD) profile out to 40kpc from the CG centre. During this fitting procedure, the central 2.5 kpc region was not included to avoid any saturated pixels. 
    \item The fitted CG profile was subtracted from the CG+IGL profile to obtain the IGL profile. Any remaining light in the central region due to fitting only beyond 2.5 kpc was excluded. 
    \item \modtexttwo{The total flux in the IGL was computed by integrating the IGL profile out to the SB limit of 30~mag/arcsec$^2$ with SNR$\geq5$ for each magnitude bin, similar to the Hydrangea stacked profiles (see Sec.~\ref{sec:sim_prediction} for details).} Similarly, the total group light was measured by integrating the total group profile out to the same radial distance. Their ratio was taken as the fraction of total light in IGL, or $f_{\mathrm{IGL}}$. 
\end{enumerate}


The measured $f_{\mathrm{IGL}}$ in the redshift range $0.09\leq z \leq 0.15$ with respect to the $\magr$ bins in our three different background subtracted images ($p_0, p_1, p_2$) are shown in Fig.~\ref{fig:f_igl_all_p_z0915}. \modtexttwo{The CG $\magr$ is taken from the GAMA \texttt{StellarMassesLambdarv20} catalogue (more details on the catalogue and measurements of magnitudes are provided in Sec.~\ref{sec:gama_data_intro})}. The shaded regions in pink, yellow, and cyan are showing the upper and lower limits of $f_{\mathrm{IGL}}$ measurements for $p_0, p_1$, and $p_2$ cases, respectively. The blue and purple dashed lines show the predictions from the Hydrangea simulations at redshifts 0.1 and 0.3, respectively (details in Sec.~\ref{sec:sim_prediction}). Our measurements of $f_{\mathrm{IGL}}$ for the $p_0$ background subtraction model is comparable to the predictions. 

\begin{figure*}
\centering
\leavevmode \hbox{%
  \includegraphics[width=0.9\textwidth]{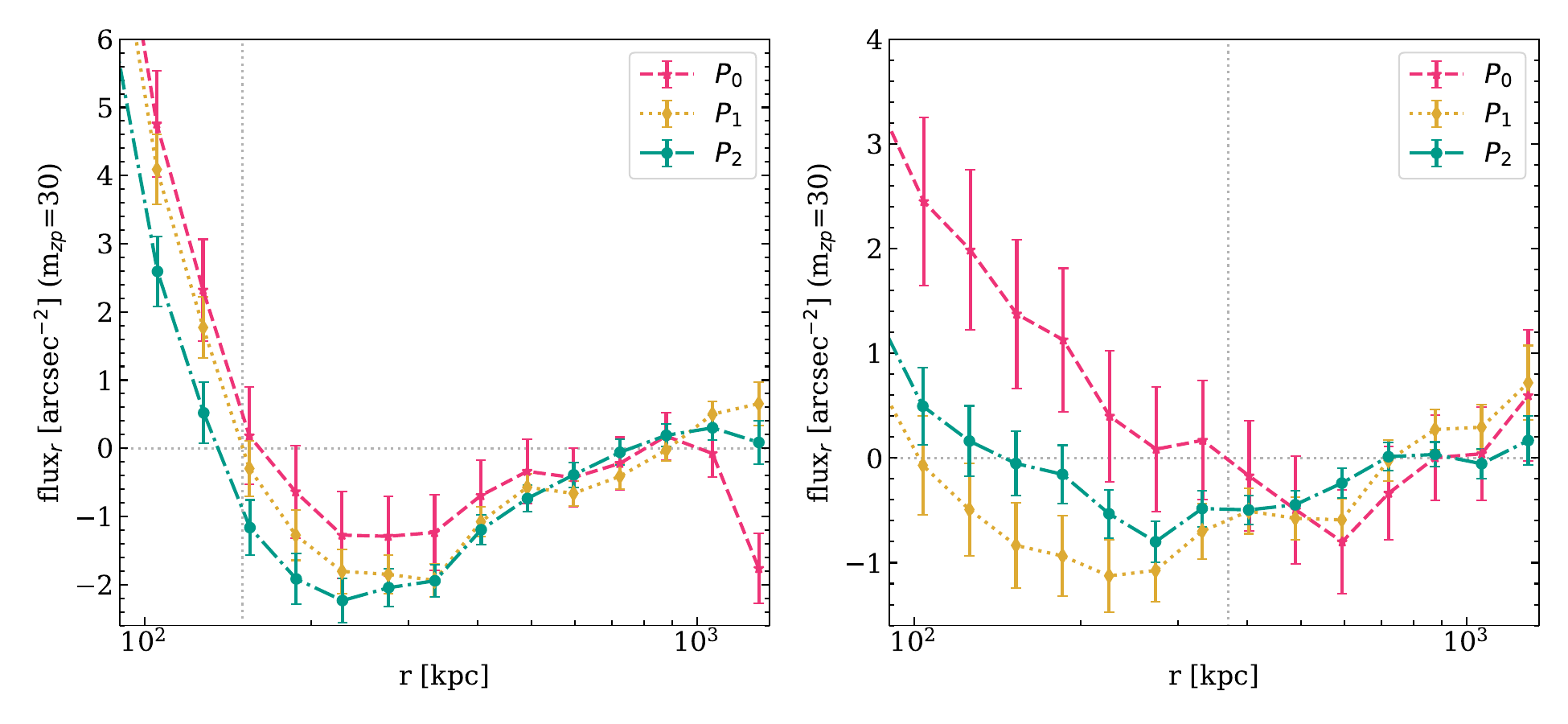}}
\caption{\modtext{Radial flux profiles of the stacked central group galaxies (CG) for the three background-subtraction models tested in this work. The left panel shows profiles of the brightest magnitude bin (mean $\magr~\approx~-23.4$ mag), and the right panel shows the faintest magnitude bin we considered at $0.09\leq z \leq 0.15$. From the central region out to 100~kpc (not shown here), their light retention is similar. For the brightest CG bin (left panel), beyond 150~kpc (vertical dotted line), the $0^{\mathrm th}$ order background-subtracted images ($p_0$) have better retention of light compared to the other two. However, considering the flattening of the profiles in the far outskirts (beyond 500~kpc) as the true sky background, the extended CG profiles between 150 and 500~kpcs are over-subtracted for all three background-subtraction models. For the faintest CG bin (right panel), the $p_0$ case shows a much better light retention compared to the brightest bin, out to 380~kpc (vertical dotted line). The $p_1$ and $p_2$ cases, 
 however, show a similar level of over-subtraction compared to the brightest bin. Both of these panels demonstrate that the $p_0$ background subtraction method has the best performance for retaining the extended faint light around the group CGs.
 \label{fig:outer_p012_profiles}}}
\end{figure*}

To define the upper and lower limit of $f_{\mathrm{IGL}}$ measurement for each background subtraction case for a specific redshift and magnitude bin (i.e. each of the shaded regions), we considered the outer regions ($\geq 200$ kpc) of the SB profiles. Figure~\ref{fig:outer_p012_profiles} shows the outer regions for all three background subtraction cases at $0.09\leq z \leq 0.15$: \modtext{the left panel shows the profiles for the brightest magnitude bin we considered (mean $\magr\approx-23.4$ mag), and the right panel shows the profiles for the faintest magnitude bin (mean $\magr\approx-22.1$ mag)}. We expect that the overall sky background always dominates beyond 500kpc from the group centres because most of the groups have a virial radius below this limit, and we see that the average flux count beyond this limit is close to zero for all $p_0, p_1$, and $p_2$ cases out to 1Mpc. However, between 150 and 500~kpc from the group centres, all three profiles show a `U' shaped down and up turn in the profile, which indicates a ring-like over-subtracted region in the image. This over-subtraction is similar for $p_1$ and $p_2$, which show a stronger over-subtraction compared to $p_0$. As there is no reliable way to recover this missing light, it poses a limiting factor in our measurements, especially for the lower redshift and brighter CG bins (left panel of Fig.~\ref{fig:outer_p012_profiles}). As described above, for each of the $p_0, p_1$, and $p_2$ cases, we fitted a horizontal line to the data points beyond 500~kpc in Fig.~\ref{fig:outer_p012_profiles} and considering it as the sky background, subtracted this value from these profiles before fitting the CG to separate the IGL (step~1 as described above). Given the over-subtracted region in the profiles, this background subtraction brings the measurable extent (before it assumes a negative value) of the SB profiles down to $\approx 190~$kpc for $p_0$, and $\approx 150~$kpc for $p_1$ and $p_2$. The IGL measurement from this background definition provided our lower limit to $f_{\mathrm{IGL}}$, which is shown in the lower bound of the shaded regions in Fig.~\ref{fig:f_igl_all_p_z0915} and Fig.~\ref{fig:f_igl_p0_all_z}. The upper limit to the measurements came from taking the minimum value of the profiles as the background value and subtracting that before the IGL measurement. 
This measurement and its scatter define the upper bounds of the shaded regions in Fig.~\ref{fig:f_igl_all_p_z0915} (and also Fig.~\ref{fig:f_igl_p0_all_z}). The $p_0$ background-subtracted profiles for all the redshift and magnitude bins presented here are shown in Appendix~\ref{sec:flux_profile_all_app}.

\subsubsection{Measurements with different background-subtraction models}

Given that we already showed in Sec.~\ref{sec:bkg_on_galaxy_profile} that the $p_0$ model has the best retention of faint light at the outskirts of the galaxy SB profiles, we conclude that our measured $f_{\mathrm{IGL}}$ for $p_0$ (pink shaded region) is the upper bound of this measurement in all the background subtraction models we considered in this work. One point of concern here is that the $p_0$ model only subtracts a constant background, and therefore can leave the most amount of residual background pattern out of the three models we used. This can particularly impact the $f_{\mathrm{IGL}}$ measurement for the fainter $\magr$ bin, causing a potential overestimation. The residual background subtraction explained in Sec.~\ref{sec:res_bkg_sub} minimizes any such overestimation. On a different note, the $f_{\mathrm{IGL}}$ measurement for the brightest $\magr$ bin is likely the best estimate out of the three background subtraction models, although even $p_0$ shows signs of some over-subtraction at the edge of the SB profile of the brightest CGs (Fig.~\ref{fig:outer_p012_profiles}). The $p_1$ and $p_2$ profiles follow each other closely, show significantly stronger over-subtraction around 200~kpc than $p_0$ (Fig.~\ref{fig:outer_p012_profiles}), and have similar $f_{\mathrm{IGL}}$ measurements (Fig.~\ref{fig:f_igl_all_p_z0915}). Considering all the above points, we decided to use only the $p_0$ background subtracted images to measure $f_{\mathrm{IGL}}$ in GAMA groups at different redshift ranges.   

\subsubsection{Impact of redshift on background and measurements}
\label{sec:z_evol_obs}

\begin{figure}
\centering
\leavevmode \hbox{%
  \includegraphics[width=0.9\columnwidth]{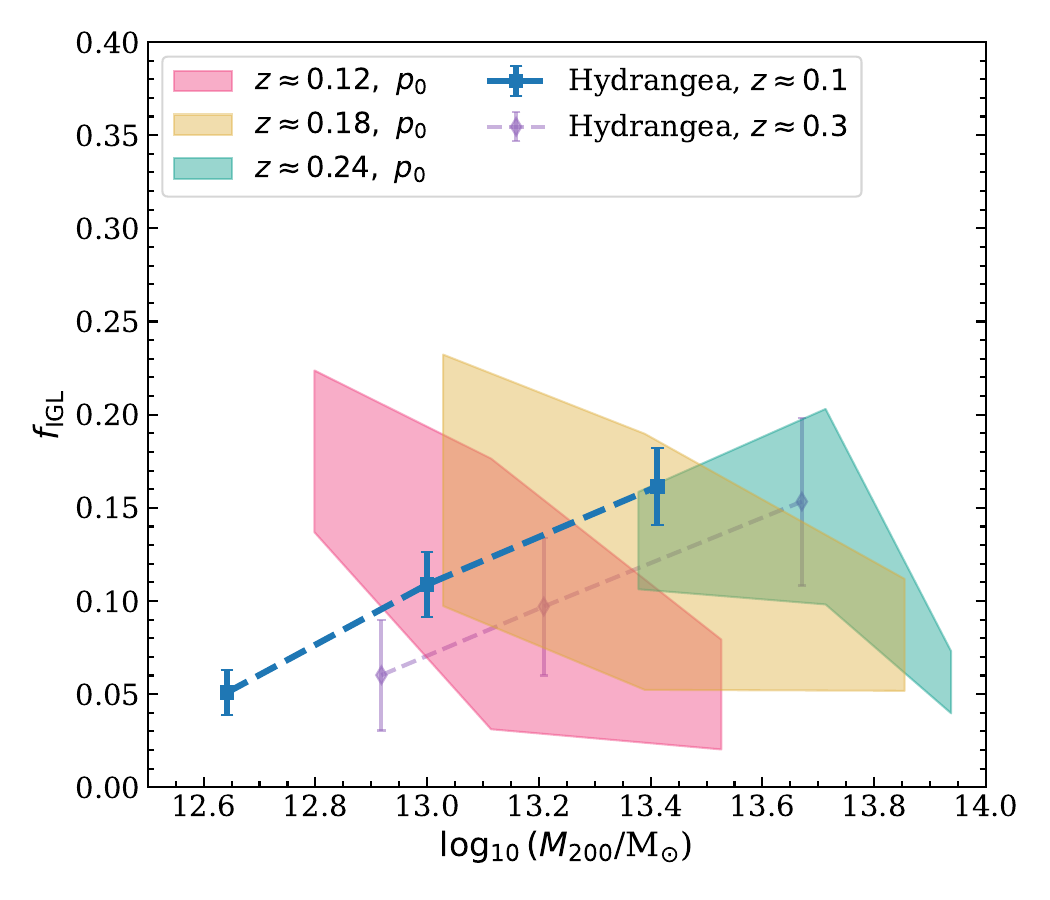}}
\caption{\modtext{Fraction of light in IGL compared to the total group light ($f_{\mathrm{IGL}}$) of the KiDS+GAMA groups at different redshifts from the $0^\mathrm{th}$ order ($p_0$) background-subtracted images. The x-axis indicates the mean halo mass $M_h$ of the corresponding bin. Pink, yellow, and cyan shaded regions show the upper and lower limit of $f_{\mathrm{IGL}}$ for the redshift ranges $0.09<z<0.15$, $0.16<z<0.21$, and $0.21,z<0.27$, respectively. Blue and purple lines show prediction of $f_{\mathrm{IGL}}$ from the Hydrangea simulations at comparable redshifts (mentioned in labels). The range of $f_{\mathrm{IGL}}$ is comparable between simulations and observations.}
 \label{fig:f_igl_p0_all_z}}
\end{figure}

Groups with the same physical size have a smaller angular size at higher redshifts. Also, for the same apparent magnitude limit of the GAMA galaxy measurements, groups with intrinsically brighter CG are more numerous at higher redshifts. Compared to the groups with similarly luminous CGs at lower redshifts, these high redshift groups have different systematics in the data due to their different angular sizes on the sky. We check the impact of these different systematics on the $f_{\mathrm{IGL}}$ measurements in three different redshift bins. The first is our lowest redshift bin, $0.09\leq z \leq 0.15$ (with an average $z\approx 0.12$), for which the $f_{\mathrm{IGL}}$ is shown in Fig.~\ref{fig:f_igl_all_p_z0915}. The other two consecutive redshift ranges are $0.16\leq z \leq0.21$ (with an average $z\approx0.18$) and $0.21\leq z\leq0.27$ (with an average $z\approx 0.24$). The $f_{\mathrm{IGL}}$ in these redshift ranges for the $0^{\mathrm th}$ order background subtracted images ($p_0$) are shown in Fig.~\ref{fig:f_igl_p0_all_z}. The shaded regions show the upper and lower limits of $f_{\mathrm{IGL}}$ at different redshift ranges as indicated in the labels (how we define these limits is discussed in Sec.~\ref{sec:f_igl_in_mr_bins}). Blue and purple lines indicate our predictions from the Hydrangea simulations at comparable redshifts. The mean halo mass of the stacked groups in each bin is shown along the $x$-axis.

Figure~\ref{fig:f_igl_p0_all_z} shows no significant redshift-evolution in the $f_{\mathrm{IGL}}$ measurements in either simulations or observations. Especially the measurements for the lowest-mass bins have similar values throughout. The measurements for the highest-mass bins slightly increase at higher redshifts, and their scatter gets smaller (as indicated by the vertical span of the shaded region).
However, these increased $f_{\mathrm{IGL}}$ values are likely not an indication of increased IGL at these slightly higher redshifts, but an improvement of the profile over-subtraction issue at extended radial distances. Because our measurements are made on cutouts extending to 1~Mpc distance from the group centres, and this distance corresponds to less than half of the angular size at $z\approx0.24$ compared to $z\approx0.12$, the residual background patterns have a smaller impact on the group SB profiles. For example, a 1~Mpc cutout can span up to 6 CCD chips in the joined field of view at $z\approx0.12$, including the uneven background at the chip edges. Compared to that, at $z\approx0.24$, a 1~Mpc cutout can span only two chips, \modtexttwo{minimizing} the large-scale residual patterns. Also, CGs with the same luminosity are fainter at higher redshifts. 
As a result, we see reduced over-subtraction compared to Fig.~\ref{fig:outer_p012_profiles}. As we measure faint and diffuse IGL, a naive initial assumption can be that the lower the redshift, the best measurement of the faint light we can get. However, after considering this background pattern issue, considering a slightly higher redshift bin can instead improve the robustness of the IGL measurement. The same reasoning motivated the IGL analysis from \citet{martinezlombilla2023} to be done on a GAMA group at $z\sim0.2$.

\subsubsection{Combined light from the CG and IGL}
\label{sec:bgg_igl_frac}

As mentioned before, the fraction of light in the IGL/ICL compared to the total light of the host system ($f_{\mathrm{CG+IGL (or ICL)}}$) varies from a few per cent to more than 30 per cent \citep[e.g.][]{Kluge2021,Montes2022}. A major contribution to this variation comes from the methods used to separate the light from the central galaxies and IGL/ICL. One way to reduce the bias introduced by the CG-IGL separation method is to simply consider the combined light from the CG and the IGL/ICL at different radial distances from the CG. As most of the light in the CG is accreted, and there is a smooth transition in the light profile where CG and IGL-dominated regions overlap, considering the CG and diffuse components together helps with standardizing the measurements of this diffuse component among different studies, as suggested by \citet{gonzalez2007}. 

\begin{figure}
\centering
\leavevmode \hbox{%
  \includegraphics[width=\columnwidth]{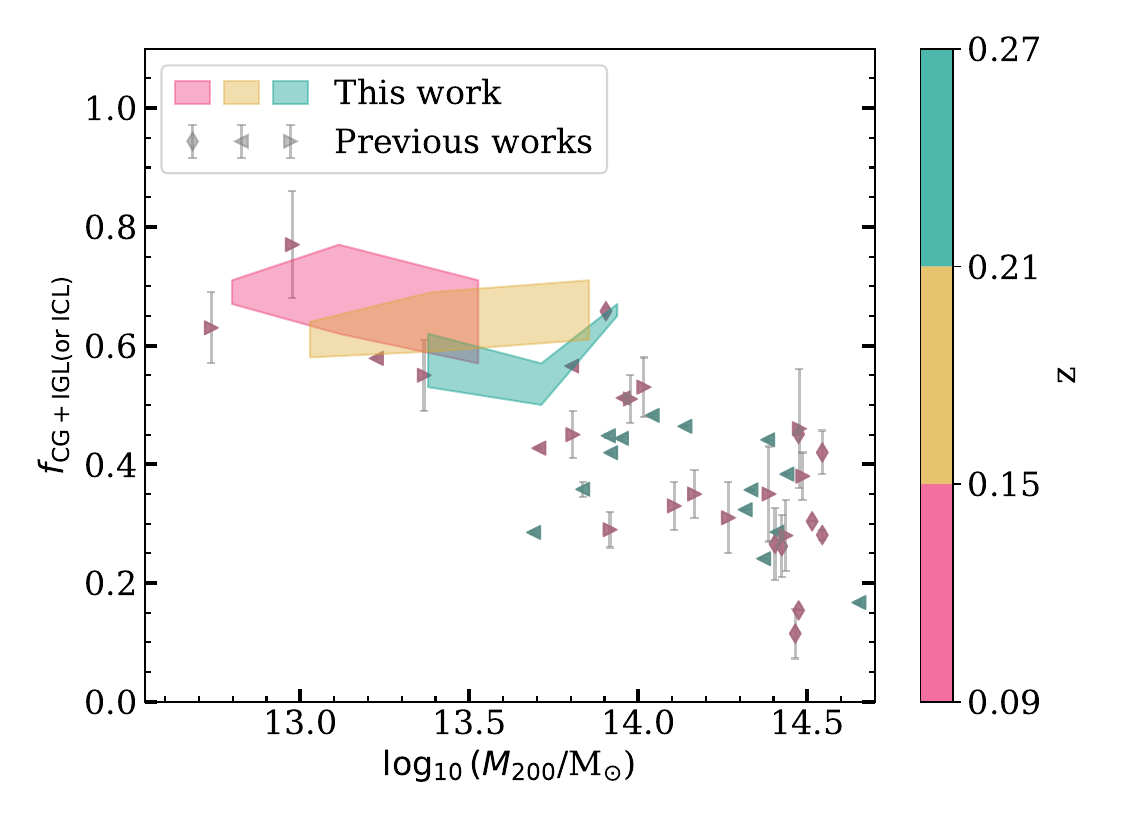}}
  \vspace{-0.5cm}
\caption{Fraction of light in CG+IGL (or ICL) compared to the total group light ($f_{\mathrm{CG+IGL (or ICL)}}$) of the KiDS+GAMA groups at different redshifts from the $0^\mathrm{th}$ order ($p_0$) background-subtracted images. The x-axis indicates the mean halo mass $M_{200}$ of the corresponding bins for this work (shaded regions), and the group/cluster halo mass for the previous works (different markers). Pink, yellow, and cyan shaded regions show the upper and lower limit of $f_{\mathrm{CG+IGL}}$ for the redshift ranges $0.09<z<0.15$, $0.16<z<0.21$, and $0.21,z<0.27$, respectively. The diamonds, left-facing-triangles, and right-facing-triangles show previous works from \citet{Kluge2021}, \citet{Furnell2021}, and \citet{gonzalez2007}, respectively. The redshift ranges of the previous works are shown in the colourbar. Our results agree with a high $f_{\mathrm{CG+IGL}}$ value for low halo masses. 
 \label{fig:f_bcg_igl_p0_all_z}}
\end{figure}

Previous works based on groups and clusters show that there is a weak negative correlation between $f_{\mathrm{CG+IGL (or ICL)}}$ and redshift, which is more likely to be driven by the observed strong negative correlation of $f_{\mathrm{CG+IGL (or ICL)}}$ and halo mass of the host group/cluster \citep{lin2004,gonzalez2005,gonzalez2007,burke2015}. 
\begin{figure}
\centering
\leavevmode \hbox{%
  \includegraphics[width=\columnwidth]{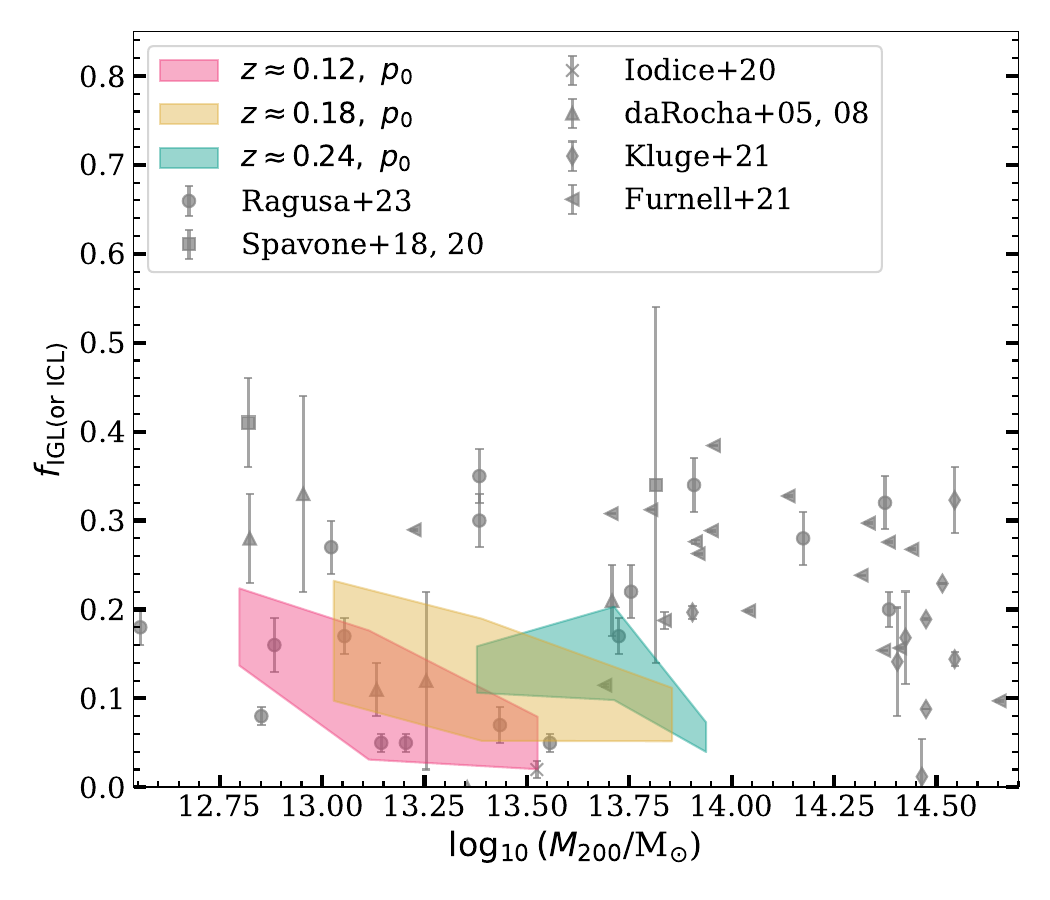}}
\caption{\modtext{Fraction of light in IGL compared to the total group light ($f_{\mathrm{IGL}}$) of the KiDS+GAMA groups at different redshifts from the $0^\mathrm{th}$ order ($p_0$) background-subtracted images. $f_{\mathrm{IGL}}$ from this work along with previous works on individual galaxy groups and clusters (grey markers). The x-axis indicates the mean halo mass $M_h$ of the host systems (previous works) or the corresponding bins (this work). Pink, yellow, and cyan shaded regions show the upper and lower limits of $f_{\mathrm{IGL}}$ for the redshift ranges $0.09<z<0.15$, $0.16<z<0.21$, and $0.21,z<0.27$, respectively.
\label{fig:f_igl_p0_all_z_prev_works}}}
\end{figure}

Figure~\ref{fig:f_bcg_igl_p0_all_z} shows the $f_{\mathrm{CG+IGL (or ICL)}}$ from this work along with previous works that explored this measurement. The colour of the shaded regions (this work) and data points (previous works) show the redshift ranges of the groups and clusters. If any data point from previous works is at $z<0.09$, it is shown in pink, and data points at $z>0.27$ are shown in cyan to show the redshift range they are closer to. Similar to the previous works, our results also demonstrate the negative correlation of $f_{\mathrm{CG+IGL (or ICL)}}$ and group halo mass, which is more visible for the lower redshift sample (pink shaded area). This negative correlation indicates that in more massive group/cluster haloes, there is more light (and therefore mass) in the satellites compared to the CG+IGL as clusters evolve by adding satellites \citep[e.g.][]{lin2004,Gonzalez2013}. Our measurements of the combined CG+IGL and IGL fractions are given in Table~\ref{tab:fparam_table}.  

\begin{table}
	\centering
	\caption{Measurements of IGL and CG+IGL in our KiDS+GAMA group sample that are shown in Fig.~\ref{fig:f_igl_p0_all_z} and Fig.~\ref{fig:f_bcg_igl_p0_all_z}.}
	\label{tab:fparam_table}
	\begin{tabular}{|l|cccr|} 
		\hline
		$z$ & $\log_{10} (\frac{M_{200}}{\msun})$ & CG Mag$_\mathrm{r}$ & L$_{\mathrm{IGL}}$/L$_{\mathrm{Tot}}$ & L$_{\mathrm{CG+IGL}}$/L$_{\mathrm{Tot}}$  
          \\  
          
		\hline
    
		~ & 12.79 & -22.40 & 0.14 - 0.19 & 0.67 - 0.71\\
		
		0.12 & 13.11 & -22.93 & 0.05 - 0.16 & 0.63 - 0.76\\
		
		~ & 13.53 & -23.43 & 0.02 - 0.07 & 0.58 - 0.70\\
		\hline
        
		~ & 13.03 & -22.33 & 0.11 - 0.21 & 0.59 - 0.63\\
		
		0.18 & 13.40 & -23.07 & 0.07 - 0.17 & 0.60 - 0.68\\
		
		~ & 13.85 & -23.62 & 0.06 - 0.10 & 0.62 - 0.70\\
		\hline
        
		~ & 13.38 & -22.60 & 0.12 - 0.15 & 0.54 - 0.61\\
		
		0.24 & 13.71 & -23.27 & 0.11 - 0.19 & 0.51 - 0.56\\
		
		~ & 13.94 & -23.81 & 0.04 - 0.07 & 0.65 - 0.67\\
  
		\hline
	\end{tabular}
    \tablefoot{First three columns show the mean redshift, halo mass, and $r-$band CG magnitudes of the CG $\magr$ bins used in this work, respectively. The value ranges in L$_{\mathrm{IGL}}$/L$_{\mathrm{Tot}}$ and L$_{\mathrm{CG+IGL}}$/L$_{\mathrm{Tot}}$ columns show the lower and upper limits of our measurements as described in Sec.~\ref{sec:f_igl_in_mr_bins}}
\end{table}

\subsubsection{Comparison to other works}
\label{sec:obs_comp}

Our measurement of the IGL fraction is comparable to other existing IGL/ICL measurements (Fig.~\ref{fig:f_igl_p0_all_z_prev_works}). Given the halo mass range of our group sample, the strength of our measurements comes from stacking many groups, which reduced the scatter in our measurements compared to existing works, especially ones that are computed in individual systems. \citet{martinezlombilla2023} measured the IGL fraction in a GAMA group with ID 4001389 (RA 35.834163 deg, DEC -5.454157; J2000) from the GAMA group catalogue G$^3$Cv10 \citep{Robotham2011} using multi-band data from the Hyper Suprime-Cam Subaru Strategic Program Public Data Release 2 \citep{aihara2019}. Their measurement of $f_{\mathrm{IGL}}$ using different surface brightness cuts and a 2D composite model spans 0.035 - 0.305 (among different methods) in the $r-$band. Although our method of separating CG from IGL is not the same one as they used, at the redshift ($z \approx0.2$) and halo mass ($M_{\mathrm{dyn}}=1.3\times10^{13}\msun$) of their measured group, our measurements indicate $\sim 0.1-0.2$ for the $p_0$ images, which is consistent with their measurements. In another recent work, \citet{ragusa2023} measured IGL/ICL fractions for VST Early-type GAlaxy Survey (VEGAS, \citealt{iodice2021}) data at $z\leq0.05$, and their $f_{\mathrm{IGL}}$ measurements for individual groups at $M_{\mathrm{vir}}<10^{14}\msun$ range between $\sim$0.2 - 0.4 (from their fig.~2). Although their measurements are from the local Universe, unlike our slightly higher redshifts, \modtext{we do not expect a lot of evolution in the IGL component over the redshifts covered in our work. The average redshift range our sample covers (between 0.12 and 0.24) corresponds to about 0.8 - 2.0 Gyr lookback time compared to the VEGAS sample. This is not enough time for the groups and IGL to evolve significantly, especially since we are measuring the stacked light in this work.} Our measurements are also similar to the IGL fraction of individual groups at different redshifts from fig.~2 of \citet{Montes2022}, groups of different halo masses (using N-body simulations) from \citet{Rudick2011}, and stacked measurements of 687 SDSS groups at $0.2\leq z \leq 0.3$ from \citet{zibetti2005}. We found in \citet{ahad2023} that the IGL fraction increases with the host halo mass and luminosity of the group CG. We do not see that trend in our KiDS+GAMA IGL measurements (as shown in Fig.~\ref{fig:f_igl_p0_all_z_prev_works}). \modtexttwo{\citet{ragusa2023} also reported the lack of this IGL-fraction to halo mass trend in their sample}. However, the range of values is comparable, and given the strong systematics in the background of our data for the brighter CGs, it is not possible to comment on the overall trend with certainty. 

As for the scatter of IGL fractions, Fig.~\ref{fig:f_igl_p0_all_z_prev_works} shows that several previous works (grey points) at comparable halo-masses found a higher fraction of IGL compared to our stacked measurements, while the rest are consistent with the stacked IGL fractions. This variation could be partly explained by the intrinsic scatter of IGL fractions in a diverse sample of galaxy groups (e.g., shown by fig.~7 and fig.~8 from \citet{ahad2023}). Another possibility is that the higher IGL fractions of previous works result from a biased sample of groups in those works. It has been shown that compact groups show a higher IGL fraction \citep[e.g.][grey upward triangles in Fig.~\ref{fig:f_igl_p0_all_z_prev_works}]{darocha2008}, and they are easier to observe. An additional reason for the scatter could be the presence of groups with different dynamical states, because relaxed and more evolved groups and clusters have been shown to have higher ICL fractions compared to less-evolved ones in both simulations and observations \citep[e.g.][]{darocha2008,Montes2018,Poliakov2021,ragusa2023,contreras-santos2024}. A systematic study of the impact of the above-mentioned factors in a stacking analysis of IGL/ICL requires a larger dataset with more detailed information on the systems. While this is out of scope for this work, this will be possible with larger group and cluster samples from the next generation of wide-field surveys such as \emph{Euclid} \citep{mellier2025} and LSST \citep{ivezi2019,brough2020}. 

Our results for the combined CG+IGL light are comparable to the previous works at comparable halo mass, as shown by Fig.~\ref{fig:f_bcg_igl_p0_all_z}. For our highest considered redshift sample ($0.21<z<0.27$, cyan), the values of $f_{\mathrm{CG+IGL (or\ ICL)}}$ are slightly higher than the median of the previous works at comparable redshifts and host halo mass, but not out of range considering the uncertainties. One reason for this higher value of $f_{\mathrm{CG+IGL (or ICL)}}$ could be the limitation in our data, as explained before. Future work on higher mass clusters using the same methodology will be needed to check this further. 


\section{Discussion and Conclusions}
\label{sec:conclusions}
\subsection{Discussion}
The IGL/ICL is an excellent probe to understand the growth of large scale structures like galaxy groups and clusters. However, due to the low surface brightness (LSB) nature of IGL/ICL, their detection and implications on structure formation are non-trivial. Stacking images of multiple groups and clusters provides a solution to improve the SNR of the IGL/ICL signal, and deep data from cosmology surveys such as KiDS is a valuable resource for such stacking analysis. However, the data processing pipeline of such surveys needs to be adjusted to retain the LSB features. In this work, we presented an updated pipeline to retain the LSB light in the deep $r-$band images from the KiDS survey, and tested its performance in measuring the IGL in galaxy groups from the GAMA group catalogue. 

Due to the persistent presence of background patterns on small and large scales in the KiDS data, even after our re-processing, it is challenging to obtain a clean measurement of the IGL. Nevertheless, using our custom background estimation and subtraction pipeline, careful selection and binning of sample groups, and stacking many groups to improve the SNR has allowed us to obtain a good constraint on the IGL measurement. This is the first well-constrained stacked measurement based on such a large sample of groups in the halo mass range we considered ($12.5\leq \log_{10}[M_{200}/\msun] \leq 14.0$). Moreover, this analysis highlights the potential of wide-field surveys for LSB analyses, such as the IGL measurement. 

A crucial factor for robust IGL (or any LSB) measurement is a uniform and flat sky background with the minimum possible residual background pattern. Wide-field cameras, by construction, pick up light from a wide area of the sky in each field-of-view, and are therefore particularly susceptible to stray light from passing objects outside of the field-of-view which cause internal reflection in the camera and non-uniform illumination patterns in the joined image. Image regions can also contain scattered light shadows from bond wire baffles. These issues need to be carefully resolved during the data processing stages. One added layer of complexity came with using the KiDS + GAMA overlapping fields-of-view. Because these overlapping regions were prioritized during the KiDS survey design, the data we used are from the early stages of the survey, which suffered from a problem with the baffling, resulting in more stray light. The later data releases have these issues resolved, but the KiDS+GAMA fields were already in place by then. Therefore, images from the later KiDS observations may be more suitable for LSB measurements once reprocessed by our custom pipeline with updated background subtraction. However, we lack a reliable group catalogue (such as the GAMA catalogue) in those regions.

Despite the challenging image data, with our carefully designed and tested analysis, we present an IGL measurement from the largest group sample to date, demonstrating the strength of a stacking analysis. Our complete pipeline (from data processing to sample selection and analysis steps) will be a useful tool for statistical analysis of the IGL across a wide halo-mass and redshift range when data from the next generation of wide-field surveys such as \emph{Euclid} and LSST are available. 

\subsection{Conclusions}
Our main findings from this work are listed below.

\begin{itemize}
    \item To optimize cosmology survey data for low-surface-brightness (LSB) analysis, the most important adjustment is to ensure a flat sky background. A non-uniform background can be caused by non-uniform illumination from reflections of stray light into the wide-field camera. On the data processing side, background patterns from over-subtraction of faint light around bright sources can be caused by local sky detection and subtraction based on small sections of area within the large image. The issues caused by the instrument can be mostly modelled and corrected, and an updated pipeline is required to resolve the background subtraction issue. We perform these before our analysis.

    \item We tested the performance of different background estimation models based on their resulting scatter of mean background values at random points in the field-of-view (Fig.~\ref{fig:bkg_m012}), and the retention of faint light in the extended galaxy profiles (Fig.~\ref{fig:bkg_p012_on_sb}). Based on these two criteria, a first-order polynomial ($p_1$) for the background model had the best performance. However, further analysis with images from all three cases showed that $p_0$ performs best in retaining the extended faint light for our brightest CG bins, and therefore, is the best background model for our study. 

    \item A comparison of the standard KiDS PSF and our updated image PSF shows improvement of faint light retention at the extended profile (Fig.~\ref{fig:psf_global}). The effect of PSF convolution on the IGL fraction ($f_{\mathrm{IGL}}$) measurement is small. 

    \item Even after the updated background subtraction, there are residual patterns in the background that affect the extended galaxy profiles where IGL dominates. Therefore it is necessary to compute and account for the residual background pattern at large radii for all the updated background-subtracted images.

    \item Not all the SB profiles of group central galaxies (CG) were usable for the stacked analysis due to the presence of significant irregularities, which could potentially bias the measurements. These irregularities were primarily caused by the presence of nearby bright sources. A representative subsample of randomly selected SB profiles (25 per cent) was visually inspected to add flags for such irregularities in the entire sample. Our conservative selection criteria resulted in leaving about half of our initial group samples in the final measurement, but it made the measurement more reliable. Because we started with a large sample, even after such a strict selection process, we had at least $\sim250 (\times5$ exposures) groups in each redshift bin we considered.

    \item We obtained upper and lower limits for $f_{\mathrm{IGL}}$ for our group sample in the lowest redshift bin ($0.09\leq z\leq 0.15$) from the three background-subtraction methods we used (Fig.~\ref{fig:f_igl_all_p_z0915}). Due to the over-subtraction of faint light at large radii, the $p_1$ and $p_2$ models provide a lower limit to $f_{\mathrm{IGL}}$, while the $p_0$ model provides an upper limit. Although the trend of $f_{\mathrm{IGL}}$ against the luminosity of group CGs from the KiDS+GAMA sample is not the same as predictions from simulations, the values are comparable. 

    \item We repeated our analysis for two higher redshift bins, with average group redshifts of $z\approx0.18$ and $z\approx0.24$ using the $p_0$ images to check for any redshift evolution in the measurement. Simulations predict a mild evolution from $z=0.3$ to 0.1. However, $f_{\mathrm{IGL}}$ from GAMA groups does not show a clear trend with redshift (Fig.~\ref{fig:f_igl_p0_all_z}). Overall, our measurements are consistent with existing works on systems with comparable halo mass and redshifts and have a smaller scatter in the measurements because of stacking many groups (Fig.~\ref{fig:f_bcg_igl_p0_all_z}, \ref{fig:f_igl_p0_all_z_prev_works}).

    \item Stacked SB profiles at higher redshifts suffer less from over-subtraction than at $z\leq0.15$, and therefore allow a more reliable measurement. This is because the same physical size of groups corresponds to a smaller angular size at higher redshifts, which spans across fewer chips at higher redshifts, consequently avoiding large-scale residual background patterns in the image mosaic. This is especially prominent for the brightest CG bin (i.e. the most massive groups) at each redshift. Therefore, in this work, at the same average halo mass (or CG luminosity) of the stacked groups, measurements from a slightly higher redshift bin are more reliable than those at a lower redshift.
  
\end{itemize}

\begin{acknowledgements}
\modtext{We thank the reviewer for valuable comments that helped to improve the clarity and presentation of this work.} The authors acknowledge support from the Netherlands Organization for Scientific Research (NWO) under Vici grant number 639.043.512 (SLA, HH). YMB acknowledges support from UK Research and Innovation through a Future Leaders Fellowship (grant agreement MR/X035166/1) and financial support from the Swiss National Science Foundation (SNSF) under project 200021\_213076. We thank M.~Kluge for providing the ICL and BCG+ICL measurements from \citet{Kluge2021}, and S.~van der Jagt for sharing the PSF measurements of the updated KiDS images.  

This research made use of data from the Galaxy and Mass Assembly survey (GAMA). GAMA is a joint European-Australasian project based around a spectroscopic campaign using the Anglo-Australian Telescope. The GAMA input catalogue is based on data taken from the Sloan Digital Sky Survey and the UKIRT Infrared Deep Sky Survey. Complementary imaging of the GAMA regions is being obtained by a number of independent survey programs including GALEX MIS, VST KiDS, VISTA VIKING, WISE, Herschel-ATLAS, GMRT, and ASKAP providing UV to radio coverage. GAMA is funded by the STFC (UK), the ARC (Australia), the AAO, and the participating institutions. The GAMA website is \href{http://www.gama-survey.org}{http://www.gama-survey.org}.

The KiDS imaging data is based on observations made with ESO Telescopes at the La Silla Paranal Observatory under programme IDs 177.A-3016, 177.A-3017, 177.A-3018 and 179.A-2004, and on data products produced by the KiDS consortium. The KiDS production team acknowledges support from: Deutsche Forschungsgemeinschaft, ERC, NOVA and NWO-M grants; Target; the University of Padova, and the University Federico II (Naples).

The analysis of this work was done using Python (\href{http://www.python.org}{http://www.python.org}), including the packages \textsc{NumPy} \citep{harris2020}, \textsc{AstroPy} \citep{astropy2013}, and \textsc{SciPy}
\citep{jones2009}. Plots have been produced with \textsc{Matplotlib} \citep{hunter2007matplotlib}. 

\end{acknowledgements}

\section*{Data Availability}
 
The data presented in the figures are available upon request from the corresponding author. The Hydrangea data are available at \href{https://ftp.strw.leidenuniv.nl/bahe/Hydrangea/}{https://ftp.strw.leidenuniv.nl/bahe/Hydrangea/}. The KiDS DR4 data are available at \href{https://kids.strw.leidenuniv.nl/DR4/access.php}{https://kids.strw.leidenuniv.nl/DR4/access.php}, and the GAMA catalogues can be accessed from \href{http://www.gama-survey.org/dr3/schema/}{http://www.gama-survey.org/dr3/schema/}.

%
%
\bibliographystyle{aa} 
\bibliography{IGL_gama_kids} 

\begin{appendix}

\section{Radial flux profiles of the sample}
\label{sec:flux_profile_all_app}

\begin{figure*}[b]
\centering
\leavevmode \hbox{%
  \includegraphics[width=\textwidth]{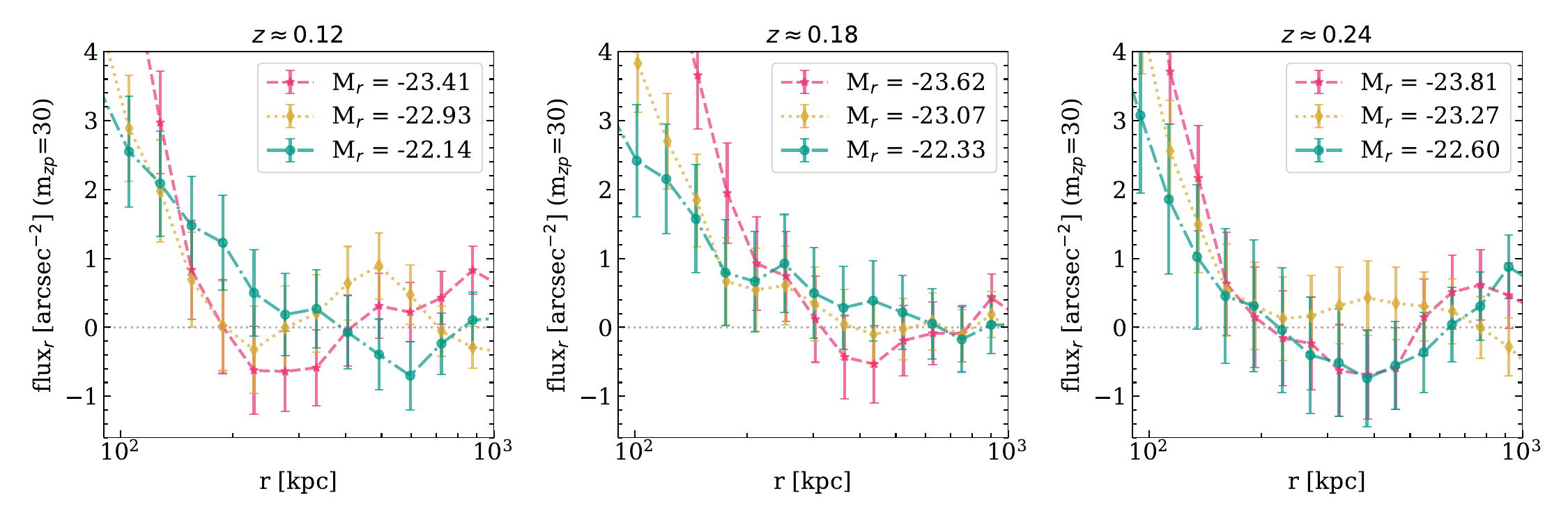}}
\caption{\modtext{Radial flux profiles of the stacked central group galaxies for all the redshift and magnitude bins in this work. The mean redshift of each subset is shown in the title of the panel, and the mean magnitude of the bins in each redshift is shown in the labels. The brightest bin is shown in red dashed line, the intermediate magnitude bin is shown in yellow dotted line, and the faintest bin is shown in green dash-dotted line. To highlight the light retention in the outer region, the profiles are zoomed-in beyond 100~kpc from the galaxy centre.
 \label{fig:all_z_mr_profiles}}}
\end{figure*}

Figure~\ref{fig:all_z_mr_profiles} shows the outer ($\geq 100$ kpc) region of the stacked radial flux profiles for all our redshift and CG magnitude bins in this work. All the profiles are from the $0^{th}$ order background subtraction procedure ($p_0$), and are adjusted for the residual background around them, indicated by the average flux count of the profiles beyond 500~kpc. These profiles show the reduced impact of oversubtraction compared to the other background subtraction methods ($p_1$ and $p_2$) shown in Fig.~\ref{fig:outer_p012_profiles}.

\modtexttwo{Even within the $p_0$ cases shown here, the impact of oversubtraction varies across different redshifts and magnitude bins, with increased oversubtraction for lower redshifts and brighter CG bins. However, the faintest bin (average $\magr=-22.6$) at $z\approx0.24$ shows worse oversubtraction compared to the lower redshifts. A few factors may be contributing to this. First, the faintest bin at $z\approx0.24$ is still brighter than the faintest bins in lower redshifts, therefore increasing the chance of oversubtraction. Another reason could be the background profiles measured at random points that were subtracted at the corresponding redshifts. At the same physical scale, the radial profiles are measured at consecutively smaller angular scales, hence increasing the uncertainty in the measurement. As the signal has a small value, even a slightly higher value of background can result in an oversubtraction. Considering the errorbars, the values are still consistent around zero.}
\end{appendix}

\end{document}